\documentstyle[12pt]{JHEP3}
\newcommand{\bej}[1]{ \begin{equation}\label{#1} }
\newcommand{\eej}{\end{equation}}
\newcommand{\beaj}[1]{\begin{eqnarray}\label{#1} }
\newcommand{\eeaj}{\end{eqnarray}}

\def\ZZZ{{\hskip-3pt\hbox{ Z\kern-1.6mm Z}}}
\def\zzz{{\hskip-3pt\hbox{ z\kern-1mm z}}}

\def\pint{{-\!\!\!\!\!\!\int}}

\newcommand{\be}{\begin{equation}}
\newcommand{\ee}{\end{equation}}
\newcommand{\ben}{\begin{eqnarray}\displaystyle}
\newcommand{\een}{\end{eqnarray}}

\def\one{{\hbox{ 1\kern-.8mm l}}}
\def\zero{{\hbox{ 0\kern-1.5mm 0}}}

\title{Classical integrability  in the BTZ black hole}
\author{Justin R. David  
and Abhishake Sadhukhan \\
 Centre for High Energy Physics,\\
Indian Institute of Science,\\ C.V. Raman Avenue, Bangalore 560012, India. \\
\email{justin@cts.iisc.ernet.in}\\
\email{abhishake@cts.iisc.ernet.in}
}

\abstract{
Using the fact the BTZ black hole is a quotient of $AdS_3$
 we show that classical string propagation in the BTZ background is
integrable. We construct the flat connection and its  monodromy matrix which 
generates the non-local charges. 
From examining the general 
behaviour of the eigen values of the monodromy matrix
we determine the set of integral equations which constrain them. 
These  equations imply
that each classical solution is characterized by  a density function in the
complex plane. For classical solutions which correspond to 
 geodesics and winding strings we solve for the 
eigen values of the monodromy 
matrix explicitly and show that   geodesics correspond to 
zero density in the complex plane.  We solve the integral  equations for
BMN and magnon like solutions and obtain their dispersion relation. 
 Finally we show that the  set of  integral equations  which constrain the eigen values of the monodromy matrix
 can be identified with  the  continuum limit of the Bethe equations of a twisted
  $SL(2, R)$ spin chain at one loop.
}

\begin{document}

\section{Introduction}

Studying the  behaviour of probes in black holes backgrounds have always revealed 
useful information about the nature of black holes.  
The geometry of the black hole 
is usually understood by studying particle motion or 
geodesics.  
 The spectrum of  scalars,  vectors, graviton  modes in the black hole background
 have also revealed useful information. 
For example for the case of black holes  which asymptotes to 
anti-de Sitter space, the study of the first  quasi-normal modes has 
revealed information  about the  transport properties of the dual field theory
\cite{Berti:2009kk}. 
In certain  cases it is also possible to study string propagation in 
black holes exactly.  These arise when the black hole background is an 
exact conformal field theory. 
A well studied example of such 
a background is the case  of the  2d  black hole  found in 
\cite{Mandal:1991tz,Witten:1991yr}.
 String propagation and its implications
 in this background  were studied  in \cite{Dijkgraaf:1991ba}.  

Instances of black hole solutions in which string propagation can be exactly solved 
are rare in higher dimensions.  3 dimensional 
Anti-de Sitter  gravity can be formulated as  a Chern-Simons theory and is topological. 
One might expect that spectrum of various excitations  about  black holes in this 
theory might be exactly solvable. Indeed  the quasi-normal modes for scalars, 
spinors and vectors in 
the BTZ black hole has been obtained analytically  in \cite{Birmingham:2001pj}. 
In fact  the BTZ black hole background along with the WZW term in the 
sigma model is an exact conformal field theory. The spectrum of strings 
in this background has been studied in the Euclidean as well as the 
Lorentzian case in \cite{Maldacena:2000kv,Troost:2002wk,Satoh:2002nj}
Thus apart from black holes in 2d gravity, the BTZ black hole offers another  
background  in which the spectrum of strings can perhaps be understood 
analytically.  In this paper we study the properties of classical strings in the 
Lorentizian BTZ background without the WZW term 
and observe that classical propagation of strings  in this 
background is integrable.  One can construct an infinite set of non-local conserved
charges. These backgrounds are quite generic in string theory, they 
occur in the near horizon limit  of the  D1-D5  black hole.
It is possible that  this structure could help in understanding  the spectrum of 
strings in this background just as in  the case of $AdS_5\times S^5$,
see \cite{Beisert:2010jr,Serban:2010sr} for a comprehensive review.

$AdS_3$ is a $SL(2, R)$  group manifold, string propagation in this background is 
classically integrable. 
One can easily show that there exists a flat connection whose 
monodromy can be used to construct a generating function of the infinite 
set of non-local charges \cite{Mandal:2002fs,Bena:2003wd}. Recently integrability along with symmetries of 
$AdS_3\times S^3$ has been used to obtain both the dispersion relation 
as well as the S-matrix of magnons for the D1-D5 system 
\cite{David:2008yk,Babichenko:2009dk,David:2010yg}. 
 The BTZ black  hole can be obtained as a quotient of $AdS_3$
 \cite{Banados:1992gq}. 
Since flatness of a connection is a local concept, the flat connection 
constructed for the case of $AdS_3$ can be used for the case of 
the BTZ black hole. However in the construction of the charges or the 
monodromy one has to impose the quotienting which makes it a BTZ black hole. 
The property that the sigma model  admits a  flat connection is probably  not true 
for black holes in higher dimensional 
Anti-de Sitter spaces. In fact recently it has been shown
 that string world sheet theory on the  $AdS_5$ Schwarzschild black hole, $AdS_5$ soliton 
as well as the on $AdS_5\times T^{1,1}$ exhibit chaos \cite{Basu:2011dg,Basu:2011di}. 
In this respect integrability of the world sheet theory for the BTZ black hole seems to be an exception. 

As a warm up exercise in studying integrability of the world sheet theory 
in the BTZ background, we  first examine the sigma model on Lens space. 
Lens space is obtained as a quotient of $S^3$,  since 
the case of sigma model on  $S^3\times R$ has been  studied in detail in 
\cite{Kazakov:2004qf}  it  is a good starting point. 
Using the existence of the  flat connection of the sigma model on $S^3\times R$,
we construct the corresponding  monodromy matrix  for the Lens space implementing 
the required quotient.  
We then study the properties of the 
quasi-momentum constructed from the monodromy matrix
 in the spectral plane and write down the set of integral equations which 
 constrain the quasi-momentum. 
  To cross check our calculations we compare the solutions obtained 
 from examining these integral equations with an explicit solution of the 
 sigma model and obtain agreement. 
 We then show that the integral equations which determine the quasi-momentum 
 are the agree with the continuum limit of the 
  Bethe equations of the BDS  long-range spin chain \cite{Beisert:2004hm}
 with an appropriate twist up to two loops.
 Quotients of $S^3$ were studied earlier \cite{Ideguchi:2004wm} 
 in the context of spin chain description for
 orbifolds of ${\cal N}=4$ super Yang-Mills but as far as we are aware the 
 twisted version of the integral equations for the sigma model on Lens space
 has not been obtained before. 
 
 We then approach the main topic of the paper. 
 Using the flat connection inherited from $AdS_3$ 
 we implement the quotient  required to obtain the 
 monodromy matrix and the quasi-momentum for classical solutions 
 of the sigma model in the BTZ$\times S^1$ background. In the main 
 text we restrict ourselves to the non-extremal BTZ black hole. 
 Most string backgrounds which contain a BTZ black hole
 also contains at least a single $S^1$ thus we do not loose
 generality. 
 The trace of the monodromy 
 matrix is the generating function of the non-local conserved charges
 of the world sheet theory.  
 We then study the properties of the quasi-momentum
 on the spectral plane  and obtain
 the integral equations which determine the quasi-momentum. 
 This allows the classification of all solutions of the sigma model on 
 the BTZ black hole in terms of the behaviour of a density defined
 on the spectral plane.  We then verify these equations 
 using two explicit class of  solutions,  geodesics  and 
 winding strings. We solve for the quasi-momentum of 
 these solutions explicitly and show that it is in agreement 
 with the properties of the quasi-momentum determined 
 from general considerations. We see that geodesics correspond to 
 zero density in the spectral plane. 
 Thus a classical solution can be described by a density function 
 on the spectral plane which satisfies a set of integral equations. 
 This feature is reminiscent of the identification of the $1/2$ BPS solutions
 in $AdS_5\times S^5$ with a density on the complex plane 
 found by \cite{Lin:2004nb}. 
 We then solve for these equations for two cases: the case of 
 a density distribution which are localized delta functions 
 and the situation in which the density is uniform and localized
 on a line. These two situations correspond to  plane wave solutions
 and  magnon solutions for the case of $S^3$. 
 Using this method we obtain dispersion relations satisfied by the 
 plane wave like solution and the magnon like solutions 
 in the BTZ background. 
 Finally we show that the integral equations satisfied by the 
 quasi-momentum are same as the Bethe equations  of a twisted 
 version of the long range $SL(2, R)$ spin chain at one loop. 
 In appendix A we study the behaviour of the quasi-momentum for the 
 case of the extremal BTZ black hole. 
 
 The organization of the paper is as follows: In the next section we study the 
 case of the Lens space as a warm up exercise. 
 Section 3 contains the main results of this paper. 
  In this section we show that 
 the world sheet theory of the sigma model on $BTZ\times S^1$ is integrable 
 and obtain the generator of the non-local charges. 
 We obtain the general properties of the quasi-momentum and verify this 
 by explicitly solving for the quasi-momentum for two classes of solutions. 
 We recast the equations satisfied by the quasi-momentum 
 in terms of  a density and show 
 that to the leading order these are same as that of the Bethe equations of a twisted 
 version of the $SL(2, R)$ spin chain. We also obtain the plane wave 
 and magnon like dispersion relations for the case of localized delta function 
 density and constant density localized on a segment respectively.  
 Section 4 contains our conclusions. 
 Appendix A  discusses  the properties of the quasi-momentum 
 for the case of the extremal BTZ black hole. 
 Appendix B has details of the two classes of solutions
 for which we explicitly obtain the quasi-momentum. 
 
 \section{Warm up example: Lens space}

To set our notations and conventions 
we first review the construction of the flat connection  and the 
monodromy matrix for the  case of $S^3$. 
The string sigma model on $R\times S^3$ can be written as 
\begin{equation}
\label{sigact1}
 S = -  \frac{\sqrt{\lambda} }{4\pi}\int d\sigma d\tau
 \left(  {\rm Tr} 
\frac{1}{2} ( g^{-1} \partial_{a} g g ^{-1} \partial^a g )  + \partial_a X_{0}\partial^a X_0
\right), 
\end{equation}
where $a = 0, 1 = \tau, \sigma$. 
The light cone coordinates are defined as 
\begin{equation}
\sigma_{\pm} = \frac{1}{2} ( \tau \pm \sigma), \qquad
\partial_{\pm} = \partial_\tau \pm \partial_\sigma.
\end{equation}
$g$ refers to the $SU(2)$ group element which we parametrize as 
\begin{equation}
\label{def-g}
g = \left(
\begin{array}{cc}
 \cos\theta e^{- i\phi} & i \sin\theta e^{i\psi} \\
i  \sin\theta e^{-i\psi} & \cos\theta e^{i\phi} 
\end{array}
\right),
\end{equation}
and $X_0$ is the time coordinate. 
Note that in this parametrization the sigma model action in (\ref{sigact1}) reduces to 
the action in the following metric of $S^3$
\begin{equation}
 ds^2 = d\theta^2  + \cos^2 \theta d\phi^2 + \sin^2 \theta d \psi^2.
\end{equation}
It is convenient to introduce the currents
\begin{equation}
\label{d-current}
j_a = g^{-1} \partial_a g,
\end{equation}
then the equations of motion  are given by
\begin{equation}
\label{eom1}
\partial_+ j_- + \partial_- j_+ = 0.
\end{equation}
Using the definition of the current given in (\ref{d-current})  it is easy to 
show that the following equation also holds
\begin{equation}
\label{flat1}
\partial_+j_-  - \partial_- j_+ + [ j_+, j_ -] = 0.
\end{equation}
We can now construct  the following connection
\begin{equation}
\label{flat-c}
 J_{\pm}(x) = \frac{j_{\pm}}{ 1\mp x}, 
\end{equation}
where $x$ is a complex number.  Using the equations of motion (\ref{eom1}) and 
(\ref{flat1}) it can be  shown  that  the above  connection satisfies  the flatness condition
\begin{equation}
 \partial_+ J_- - \partial_- J_+ + [J_+, J_-] =0.
\end{equation}
The monodromy matrix is constructed from this flat connection as follows.
\begin{eqnarray}
\label{defomega}
 \Omega(x) &=& P \exp\left( - \int_{0}^{2\pi} d\sigma J_{\sigma}\right), \\ \nonumber
&=& P \exp  \left( - \frac{1}{2} \int_0^{2\pi} d\sigma \left[ \frac{j_+}{1-x} - \frac{j_-}{ 1+x} \right]\right).
\end{eqnarray}
The above path ordered integral is performed at a constant world sheet time. 
To obtain constants of motion, we need to take the trace of the monodromy matrix.  
The reason is as follows:  consider the world sheet time derivative of the monodromy matrix we obtain
\begin{eqnarray}
\label{time-der}
 \partial_\tau \Omega(x) &=&-  \int_{0}^{2\pi}  d\sigma \Omega( 2\pi, \sigma ) \partial_\tau J_{\sigma} 
\Omega( \sigma, 0) , \\ \nonumber 
&=& - \int_{0}^{2\pi} \Omega(2\pi, \sigma) ( \partial_\sigma J_\tau - J_\tau J_\sigma  +J_\sigma J_\tau)|_{\sigma} 
\Omega( \sigma, 0) , \\ \nonumber
&=&- (  J_\tau (2\pi) \Omega(2\pi, 0) - \Omega(2\pi, 0) J_\tau(0)  ), 
\end{eqnarray}
where
\begin{equation}
 \Omega(\sigma, \sigma' ) = P \exp\left(- \int_{\sigma'}^{\sigma} 
 d\hat\sigma J_\sigma ( \hat \sigma) \right).  
\end{equation}
In the above equations we have used the flatness condition of $J$. In proceeding from the 
second line to the third line in (\ref{time-der}) we have integrated by parts and used the equations
\begin{equation}
 \partial_{\sigma} \Omega(\sigma, \sigma') = - 
 J_{\sigma}  \Omega(\sigma, \sigma') , \qquad
 \partial_{\sigma'} \Omega(\sigma, \sigma') = \Omega(\sigma, \sigma') J_{\sigma'}.
\end{equation}
Therefore from the last line of  (\ref{time-der}) we see that if 
the connection $J_{\tau}$ is periodic in $\sigma$ with period $2\pi$ we find the 
the trace of the monodromy matrix 
\begin{equation}
{\rm Tr} (\Omega(x)), 
\end{equation}
is a constant of motion. 
For the case of $S^3$, since it is the  $SU(2)$ group manifold, 
the eigen values of the monodromy matrix will be of the form
$\{e^{ip(x)}, e^{-ip(x)} \}$. Thus we have 
\begin{equation}
{\rm Tr} (\Omega(x)) = 2\cos p(x), 
\end{equation}
where $p(x)$ is called the quasi-momentum and $x$ the spectral parameter.

We can now proceed to obtain the monodromy matrix for the sigma model 
on Lens space. 
Lens space is obtained from $S^3$ with the following identifications
\begin{equation}
\phi\sim \phi + \frac{2\pi }{k_1}, \qquad \psi \sim \psi+  \frac{2\pi }{k_2}
\end{equation}
Under this identification the $SU(2)$ group element $g$ given in (\ref{def-g}) is identified as 
\begin{eqnarray}
\label{suident}
 g &\sim& \tilde A_{(1,1)}  g A_{(1,1)}, \\ \nonumber
{\mbox{where}} \;\;
\tilde A_{(n_1, n_2)}  &=&  \left( 
\begin{array}{cc}
 e^{ - i \pi ( \frac{n_1 }{k_1} -  \frac{n_2}{k_2} ) } & 0 \\
0 & e^{ i \pi ( \frac{ n_1}{k_1}-   \frac{n_2} {k_2} )} 
\end{array}  
\right), \\ \nonumber {\mbox{and}} \;\;
 A_{(n_1, n_2)}  &=&  \left( 
\begin{array}{cc}
 e^{ - i \pi ( \frac{n_1 }{k_1} +  \frac{n_2}{k_2} ) } & 0 \\
0 & e^{ i \pi ( \frac{ n_1}{k_1}+   \frac{n_2} {k_2} )}
\end{array}  
\right).
\end{eqnarray}
From the above equation
 it is clear that the general boundary conditions for the currents $g^{-1} \partial g$ is given by 
\begin{equation}
\label{lensbc}
 g^{-1} \partial g( \tau, 2\pi) = A^{-1}_{(n_1, n_2)}  g^{-1} \partial g(\tau,  0 )  A_{(n_1, n_2)}, 
\end{equation}
$(n_1, n_2)$ can be thought of as winding numbers. 
From the expression for the flat connection in terms of the currents given in (\ref{flat-c}) 
 we see that the above boundary conditions induce the same periodicity conditions 
on the flat connection. We write this down explicitly for future reference 
\begin{equation}
 J_{\pm} (\tau, 2\pi ) =   A^{-1}_{(n_1, n_2)}  J_{\pm}(\tau, 0) A_{(n_1, n_2)}.
\end{equation}
To construct the monodromy matrix we need to consider the following 
\begin{eqnarray}
 \hat{\Omega}_{(n_1, n_2)}(x)  &=& A_{(n_1, n_2)} \Omega(x), \\ \nonumber
&=& 
A_{(n_1, n_2)}  P \exp\left( - \int_{0}^{2\pi} d\sigma J_{\sigma}\right) , \\ \nonumber
 &=& A_{(n_1, n_2)}  P \exp\left(  - \int_0^{2\pi} d\sigma \frac{1}{2} \left[ \frac{j_+}{1-x} - \frac{j_-}{ 1+x} 
\right]\right), 
\end{eqnarray}
where $\Omega(x)$ is defined in (\ref{defomega}). 
Going through the same steps as for the case of $S^3$, it is easy to 
demonstrate that  the ${\rm{Tr}} ( \hat{\Omega}_{(n_1, n_2)} (x) ) $ is 
independent of worldsheet time.  The quasi-momentum in this case  can be obtained by 
\begin{equation}
 2\cos  p(x) = {\rm Tr} ( \hat \Omega (x) ).  
\end{equation}
To proceed further let us define the conserved charges as follows:
Let us expand the currents in terms of Pauli matrices as
\begin{eqnarray}
 j_a = g^{-1} \partial_a g 
= \frac{1}{2i} j_a^i \sigma^i, 
\end{eqnarray}
where $\sigma^i$ are the Pauli matrices.
Similarly the left currents are
 \be
 l_{a}=gj_{a}g^{-1}= \partial_a g g^{-1}=\frac{l_{a}^{i}\sigma^{i}}{2i}. 
 \ee
The conserved charges  corresponding to the $\sigma^3$ component of these  currents are given by 
\begin{eqnarray}
\label{lensleft}
Q_{R}^{3}&=&\frac{\sqrt{\lambda}}{4\pi}\int d\sigma j_{0}^{3}=L-2J,  \\
\label{lensright}
 Q_{L}^{3}&=&\frac{\sqrt{\lambda}}{4\pi}\int d\sigma l_{0}^{3}=L. 
\end{eqnarray}
The energy of the string solution which is generated by 
global time translations  is given by
\be
\Delta=\frac{\sqrt{\lambda}}{4\pi}\int_{0}^{2\pi} d\sigma \partial_{\tau}X_{0}=\sqrt{\lambda}\kappa. 
\ee
In defining these charges we have used the notations of \cite{Kazakov:2004qf}.
Finally the Virasoro constraints are given by 
\begin{equation}
 {\rm Tr} (j_+^2) = {\rm Tr} (j_-^2) = - 2 \kappa^2. 
\end{equation}

\subsection{Properties of the quasi-momentum}

We now study the properties of the quasi-momentum in the complex $x$-plane which is 
called the spectral plane. 
From these properties we show that all classical solutions of the sigma model can be 
characterized in terms of a density on the spectral plane which satisfies 
certain integral equations.   

The quasi-momentum $p(x)$ has poles at $x\mp 1$.
This is evident form the  same the asymptotic analysis 
as done in \cite{Kazakov:2004nh}.  The quotienting by the lens space 
does not affect this behaviour. 
This is due to the fact that  this behaviour can be obtained by  an  analysis
for which the dependence of the worldsheet $\sigma$ coordinate 
can be ignored  \footnote{We will repeat this
analysis for the case of the BTZ background in section 3.}.
From this analysis 
it can be shown that the quasi-momentum has poles at $x\rightarrow \pm1$
with the residues given by 
\be
\label{mom-pole}
p(x)=-\frac{\pi \kappa}{x\pm 1}+... , \qquad (x\rightarrow \mp 1)
\ee
We now expand the quasi-momentum at zero and at infinity. 
As $x\rightarrow \infty$, we obtain 
\begin{equation}
\label{infexp}
J_{\sigma}=- \frac{j_{0}}{x}- \frac{j_{1}}{x^{2}}+ \cdots. 
\end{equation}
Substituting this expansion in the expression for the quasi-momentum we obtain 
\begin{eqnarray}
\label{lensq}
 2\cos p(x)&=&{\rm Tr}[A_{n_1, n_2}
 P \exp \int_{0}^{2\pi} d\sigma (\frac{j_{0}}{x}+\frac{j_{1}}{x^{2}} )] \\ \nonumber
&=&{\rm Tr}\left\{ \left[\cos\pi\left( \frac{n_1}{k_1} + \frac{n_2}{k_2} \right) -
i\sin\pi \left( \frac{n_1}{k_1} + \frac{n_2}{k_2}  \right)\sigma_{3}\right]\right.  \\ \nonumber 
& \times&  \left. \left[1  -\frac{i\sigma_{i}}{2x}\int_{0}^{2\pi}d\sigma j_{0}^{i} -
\frac{i\sigma_{i}}{2 x^{2}}\int_{0}^{2\pi}d\sigma j_{1}^{i} 
-\frac{\sigma_{i} \sigma_{j}}{4x^{2}}\int_{0}^{2\pi}d\sigma \int_{0}^{\sigma}d\sigma ' j_{0}^{i}(\sigma)j_{0}^{j}(\sigma ')\right]\right\}  \\ \nonumber
&=& 2 \cos \pi \left( \frac{n_1}{k_1} + \frac{n_2}{k_2}\right)-
\frac{1}{x}{\sin \pi\left( \frac{n_1}{k_1} + \frac{n_2}{k_2} \right)  } 
\int_{0}^{2\pi} d\sigma j_{0}^{3}  -\frac{ \hat {\cal C}}{x^2}  + \cdots, 
\end{eqnarray}
where
\begin{eqnarray}
\hat{\cal C} = 
&& 
\frac{1}{2}\cos\pi \left( \frac{n_1}{k_1} + \frac{n_2}{k_2}  \right)
\int_{0}^{2\pi}d\sigma \int_{0}^{\sigma}
d\sigma ' j_{0}^{i}(\sigma)j_{0}^{i}(\sigma ') \\ \nonumber
&& +\frac{1}{2}\sin \pi \left(\frac{n_1}{k_1} + \frac{n_2}{k_2} \right)
\left [\int_{0}^{2\pi} d\sigma j_{1}^{3}
+\int_{0}^{2\pi} d\sigma \int_{0}^{\sigma} d\sigma' j_{0}^{1}(\sigma) j_{0}^{2}(\sigma')
\right. \\ \nonumber
&&
\left. 
-\int_{0}^{2\pi} d\sigma \int_{0}^{\sigma} d\sigma' j_{0}^{2}(\sigma) j_{0}^{1}(\sigma')
\right].
\end{eqnarray}
In the above equation we have retained terms to $O(1/x^2)$ to show the first non-local charge
explicitly.  
The leading terms in the expansion as $x\rightarrow \infty$ can be obtained from 
 the second equality in (\ref{lensq}). Thus the quasi-momentum 
as $x\rightarrow \infty$ is given by 
\be
\label{lensinf}
p(x)=-\pi \left(\frac{n_1}{k_1} + \frac{n_2}{k_2} \right) 
 - \frac{2\pi(L-2J)}{\sqrt{\lambda} x}+...,  \qquad (x\rightarrow \infty), 
\ee
where we have used (\ref{lensleft}). 
We now examine the behaviour of the quasi-momentum as
$x\rightarrow 0$. To do this we can use the following relation
\begin{equation}
\label{zeroexp}
 \partial_{\sigma}+j_{1}+xj_{0}+ \cdots= g^{-1}(\partial_{\sigma}+xl_{0}+ \cdots )g, 
\end{equation}
in the equation for the monodromy matrix to obtain
\be
\hat \Omega(x)=A_{(n_1, n_1)} g^{-1}(2\pi)P(\exp{-x\int_{0}^{2\pi}d\sigma l_{0}+...})g(0).
\ee
This gives the following expression for the quasi-momentum
\be
2\cos p(x)={\rm Tr}\left\{ \tilde A_{(n_1, n_1)}^{-1}
P\exp\left(-x\int_{0}^{2\pi}d\sigma l_{0}+...\right)\right\}, 
\ee
where we have used the equation given in (\ref{lensbc}). 
Proceeding as before  and substituting for the charge using (\ref{lensright}) we obtain
\be
\label{lenszero}
p(x)=2\pi m
+\pi \left( \frac{n_1}{k_1} - \frac{n_2}{k_2} \right) +
\frac{2\pi L }{\sqrt{\lambda}}x+\cdots,   \qquad (x\rightarrow 0).
\ee
Finally the quasi-momentum satisfies the following jump condition 
above and below the branch cuts
\begin{equation}
\label{lensjump}
p(x+ i\epsilon) + p(x-i\epsilon) = 2\pi n_k, 
\end{equation}
where $k$ labels the branch cuts and $n_k$ is an integer. 
The $i\epsilon$ is used to denote the value of the quasi-momentum 
above and below the branch cut. 
This condition arises because the monodromy matrix is an 
unimodular matrix and the quasi-momentum above
and below the branch cuts correspond to the two different 
eigen values. The condition given in (\ref{lensjump}) just 
ensures the unimodularity of the monodromy matrix. 

We will now follow the procedure 
 discussed in \cite{Kazakov:2004qf}  to show that given the quasi-momentum satisfying the properties discussed
 above one can define a density on the spectral plane $x$, which satisfies 
 a set of  integral equations. 
 Thus each classical solution of the sigma model on $S^3$ corresponds to 
 a density defined on the spectral plane since each classical solution 
 defines a quasi-momentum. 
 To show this we first define the resolvent given by 
 \be
G(x)=p(x)+\frac{\pi \kappa}{x-1}+\frac{\pi \kappa}{x+1}+\pi(\frac{n_1}{k_1} + \frac{n_2}{k_2} ).
\ee
The above resolvent is written by removing the singularities at 
$x\rightarrow \pm 1$.  We also subtract  
$-\pi \left(\frac{n_1}{k_1} + \frac{n_2}{k_2} \right) $  so as to make
$G(x)\sim \frac{1}{x}+\cdots $, as $x\rightarrow \infty$. 
This ensures that $G(x)$ does not have a pole at $\infty$. 
Now since $G(x)$ is an analytic function 
without any poles,  by using 
standard complex analysis 
it can  be represented by an integral of density  by 
\be
G(x)=\int d\xi \frac{\rho (\xi)}{x-\xi}.
\ee
where the integral is along the branch cuts  and $\rho(x)$ is given by 
\begin{equation}
\rho(x) = \frac{1}{2\pi  i} \left( G(x+i \epsilon) - G(x-i\epsilon)\right).  
\end{equation}
The shift in $\epsilon$ is to denote the value of the resolvent above and 
below the branch cuts. 
>From the asymptotic behaviour of the resolvent at $x\rightarrow \infty$
given in (\ref{lensinf})  we obtain the following normalization conditions on the density 
\be
\label{comp1}
\int dx \rho (x)= \frac{2\pi}{\sqrt{\lambda}}(\Delta +2J-L).
\ee
Also from the behaviour of the resolvent at $x\rightarrow 0$ as given in 
the equation (\ref{lenszero}) we see that 
\begin{equation}
  G(x)=2\pi (m+\frac{n_{1}}{k_{1}})+\frac{2\pi }{\sqrt{\lambda}}( L-\Delta) x + O(x^2) , 
  \end{equation}
 we get the following equations:
\begin{eqnarray}
\label{comp2}
\frac{-1}{2\pi i}\oint \frac{G(x)dx}{x}&=&\int dx\frac{\rho(x)}{x}=-2\pi (m+ \frac{n_{1}}{k_{1}}),  \\
\label{comp3}
\frac{-1}{2\pi i}\oint \frac{G(x)dx}{x^2}&=&\int dx\frac{\rho(x)}{x^2}=\frac{2\pi (\Delta -L)} {\sqrt{\lambda}}. 
\end{eqnarray}
Finally the unimodular condition given in (\ref{lensjump}) can be
 be recast in the following integral equation for the density $\rho(x)$. 
\begin{eqnarray}
\label{compjum}
G(x+i\epsilon )+G(x-i\epsilon )&=&2\pint d\xi \frac{\rho(\xi)}{x-\xi},  \\ \nonumber
&=&
\frac{2\pi \kappa}{x-1}+\frac{2\pi \kappa}{x+1}+
2\pi (n_{k}+\frac{n_1}{k_1}+\frac{n_2}{k_2}). 
\end{eqnarray}
The integration $\pint$ means that the integration in the complex plane has been done by excluding and moving the contour around the poles.
The integral equations (\ref{comp1}), $\;\;$ (\ref{comp2}), (\ref{comp3})  and 
(\ref{compjum}) are the conditions satisfied by the density $\rho(x)$. 
Thus we have shown that given a classical solution, it defines a 
density on the spectral plane satisfying certain integral equations.

\subsection{ The rotating string}
In this section we write down the classical solution for the 
rotating string and the dispersion relation satisfied by this solution. 
We then obtain the resolvent for this solution using the 
condition discussed in the previous subsection and show that the 
dispersion relation obtained from the resolvent agrees with that 
obtained from the explicit solution. 
Thus this comparison serves as a check on the equations satisfied 
by the density obtained in the previous subsection. 
A similar analysis for the rotating string in $S^3$ was performed
in \cite{Kazakov:2004qf}. 

To write down the solution corresponding to the rotating string it is convenient to parametrize the 
group element $g$ as the following 
\begin{equation}
 g = \left( \begin{array}{cc}
      X_1 + i X_2 &  i ( X_3 + i X_4) \\
i ( X_3 - i X_4)  & X_1 - i X_2 
     \end{array}
\right),
\end{equation}
together with the constraint
\begin{equation}
 X_1^2 + X_2^2 + X_3^2 + X_4^2 = 1.
\end{equation}
The ansatz for the rotating string is given by
\begin{eqnarray}
X_{1}+iX_{2}=\cos{\frac{\theta _0}{2}{e^{-iw_{1}\tau+i\frac{m'_1}{k_1}\sigma}}}, & & \qquad X_{3}+iX_{4}=\sin{\frac{\theta _0}{2}{e^{-iw_{2}\tau+ i\frac{m'_2}{k_2}\sigma}}}, \\ \nonumber
t &=& \kappa \tau, 
\end{eqnarray}
where $m_1'$, $m_2' $ are integers.  
Note that this ansatz satisfies the twisted boundary 
condition 
\begin{equation}
 g( \tau, \sigma + 2\pi) = \tilde A_{(-m_1' ,m_2')} g(\tau, \sigma) A_{(-m_1' m_2')}. 
\end{equation}
The equations of motion reduce to the following algebraic equation
\be
\label{lenseom}
w_{1}^2-\left(\frac{m'_1}{k_1}\right)^2=w_{2}^2-\left(\frac{m'_2}{k_1}\right)^2, 
\ee
while the Virasoro constraints reduce to 
\begin{eqnarray}
\label{lensvir1}
 \kappa^2=\left(w_1^2+( \frac{m'_1}{k_1})^2\right)\cos^2{\frac{\theta_0}{2}}
+\left(w_2^2+(\frac{m'_2}{k_2})^2\right)\sin^2{\frac{\theta_0}{2}} &=& \kappa^2,   \\ 
 w_1\frac{m'_1}{k_1}\cos^2{\frac{\theta_0}{2}}+ w_2\frac{m'_2}{k_2}\sin^2{\frac{\theta_0}{2}}&=&0. 
\end{eqnarray}
The energy $\Delta$ and the  angular momentum corresponding to rotation
in the $X_1-X_2$ and $X_3-X_4$ plane  is given by
\begin{eqnarray}
\Delta = \sqrt{\lambda} \kappa , \qquad 
J_1 = \sqrt{\lambda} \cos^2{\frac{\theta_0}{2}} w_1 , \qquad
J_2= \sqrt{\lambda} \sin^2{\frac{\theta_0}{2}} w_2.
\end{eqnarray}
$J_1$ and $J_2$ are related to $Q_R^3, Q_L^3$ defined in (\ref{lensright}) , (\ref{lensleft}) 
by the following relations
\begin{equation}
 Q_R^3 = J_1 - J_2 = L- 2J, \qquad Q_L^3 = J_1 + J_2 = L. 
\end{equation}
Now using the equations of motion (\ref{lenseom}) 
 and the  second equation in (\ref{lensvir1}) we can eliminate
$\omega_2$ to obtain the following equations
\begin{eqnarray} 
\left[w_1^{2}+\left(\frac{m'_{2}}{k_2}\right)^2-
\left(\frac{m'_{1}}{k_1}\right)^2\right](J_1-\sqrt{\lambda}w_1)^2-(J_2 w_1)^2&=&0,  \\
 J_1 \frac{m'_{1}}{k_1}+J_2 \frac{m'_{2}}{k_2}&=&0. \nonumber
\end{eqnarray}
The last line is just a rewriting of the Virasoro constraints using the definition of $J_1$  and 
$J_2$. Now  a solution to these equations is the following
\begin{equation}
 \frac{m_1'}{k_1 } = -\frac{m_2'}{k_2}.
\end{equation}
For this solution we can derive a dispersion relation by using the first Virasoro constraint in 
(\ref{lensvir1}. This results in 
\be
\label{lenssol}
\Delta^2=(J_1+J_2)^2+\lambda \left| \frac{m_1'}{k_1} \frac{ {m_2}}{k_2} \right|,  
\ee
Now for large $L= J_1+J_2$ and for  
\begin{equation}
  \frac{m_1'}{k_1 } = -\frac{m_2'}{k_2} + O(1/L), 
\end{equation}
then the dispersion relation is basically given  by the leading approximation of 
(\ref{lenssol}). Thus we obtain
\be
\label{lensdisp1}
\Delta=L+ \frac{\lambda}{2L} \left| \frac{m_1'}{k_1} \frac{ {m_2}}{k_2} \right|  + O(1/L^2).
\ee
This relation was obtained for the twisted rotating string earlier  by \cite{Ideguchi:2004wm}. 

We will now obtain the dispersion relation given in (\ref{lensdisp1}) by constructing the resolvent
which solves the equations (\ref{comp1}), (\ref{comp2}), (\ref{comp3}) and (\ref{compjum}).  
Following the analysis in \cite{Kazakov:2004qf}
 we rescale $x$ by $x\rightarrow \frac{x}{4\pi \kappa}$ 
 in the equations  determining the resolvent. Thus they reduce to 
\begin{eqnarray}
\int dx \rho (x)&=& \frac{J}{\Delta}+\frac{\Delta-L}{2\Delta},  \\ \nonumber
\frac{-1}{2\pi i}\oint \frac{G(x)dx}{x}&=&- 2\pi (m+\frac{n_{1}}{k_{1}}),  \\ \nonumber
\frac{-\lambda}{8 \pi^2 \Delta}\oint \frac{G(x)dx}{2\pi i x^2}&=&\Delta -L ,  \\ \nonumber
G(x+i0)+G(x-i0)&=&2\pint d\xi \frac{\rho(\xi)}{x-\xi},  \\ \nonumber
&=&\frac{1}{2}\left(\frac{1}{x-1}+\frac{1}{x+1}\right) 
+ 2\pi \left(n+\frac{m_1}{k_1}+ \frac{m_2}{k_2}\right). 
\end{eqnarray}
Note that these equations are the same as the ones obtained by \cite{Kazakov:2004qf}
but with the following replacements \footnote{ See equations (5.39) in that paper.}  
\begin{equation}
 n\rightarrow  -( n + \frac{n_1}{k_1} + \frac{n_2}{k_2} ), \qquad
m \rightarrow -(m + \frac{n_1}{k_1}) .
\end{equation}
where the variables on the left hand side refers to the variables used in 
\cite{Kazakov:2004qf}
The form of the resolvent which satisfies these is given by 
\begin{eqnarray}
G(x)& =& \frac{1}{4}\left(\frac{1}{x-a}+\frac{1}{x+a}\right)+
\frac{1}{4}\left(\frac{(1+\epsilon)^{-1/2}}{x-a}+\frac{(1-\epsilon)^{-1/2}}{x+a}\right)\sqrt{Ax^2+Bx+C}
\nonumber \\ 
& &  +\pi \left(n+\frac{m_1}{k_1}+\frac{m_2}{k_2}\right). 
\end{eqnarray}
Following the same steps as in  \cite{Kazakov:2004qf}
it is possible to solve for the constants 
$A, B, C, \epsilon$ and obtain the dispersion relation 
\be
\Delta=L+\frac{\lambda}{2L}\left| (n-m+\frac{n_2}{k_2})(m+\frac{n_1}{k_1})\right|+...
\ee
We can now compare with the equation (\ref{lensdisp1}) and see that the anomalous
dimension has the same form. That is it is a product of 
a fractions of $m_1'/k_1$ and $m_2'/k_2$. 
This agreement of obtaining the dispersion relation from the 
actual solution as well as from the resolvent provides 
a check on the equations   (\ref{comp1}), (\ref{comp2}), (\ref{comp3}) and (\ref{compjum})
which determine the density corresponding to each classical 
solution in the Lens space.

\subsection{The relation with the $SU(2)$ spin chain}

In this section we show that the equations which determine the density 
can be obtained from the continuum limit of the Bethe equations of 
a twisted version of the long range BDS spin chain introduced by 
\cite{Beisert:2004hm}.
The twisted spin chain we consider satisfies the following Bethe equations
\begin{equation}
\label{bethe1}
\left( \frac{ x( u_k + \frac{i}{2} )}{ x(u_k - \frac{i}{2} )} 
\right)^{L}  \exp\left[2 \pi i \left(  \frac{n_1}{k_1} + \frac{n_2}{k_2}  \right)\right] 
= \prod_{j= 1, j\neq k}^{ J } \frac{ u_k - u_j + i}{ u_k - u_j - i}, 
\end{equation}
where $x$ as a function of $u$ is given by 
\begin{equation}
\label{chgvarxu}
x(u) = \frac{u}{2} + \frac{u}{2} \sqrt{1 - \frac{ 2g^2}{u^2} }, 
\qquad u(x) = x + \frac{g^2}{2x} . 
\end{equation}
The $x$'s are the Bethe roots and related to the momentum of the spin wave excitations of
the chain. 
The cylicity constraint is given by
\begin{equation}
\prod_{k=1}^{ J}\left( \frac{ x( u_k + \frac{i}{2} )}{ x(u_k - \frac{i}{2} )} 
\right) =  \exp\left(- 2\pi i  ( m + \frac{n_2}{k_2} ) \right), 
\end{equation}
and the energy of the spin chain is given by 
\begin{equation}
D = 2g^2 \sum_{i =1}^J \left( \frac{i}{x( u+ \frac{i}{2} )} - \frac{i}{x( u- \frac{i}{2} ) }
\right). 
\end{equation}
These equations define the  twisted version of the long 
range $SU(2)$ chain. On setting the twists $n_1=n_2=0$, the twisted version 
of the spin chain reduces to that studied in \cite{Beisert:2004hm}. 
We will show that the continuum limit of these equations reduces to that 
satisfied by resolvent of the sigma model. 

To take the continuum limit we perform the following scaling $u_k \rightarrow L u_k$
with $L \rightarrow\infty$. After
taking logarithm on both sides of the equation (\ref{bethe1}) we obtain
\begin{equation}
\label{bethe2}
 L \ln \left( \frac{ 1+ \frac{i}{2L u_k} }{ 1- \frac{i}{2Lu_k} } \right)
+ 2\pi i \left( \frac{n_1}{k_1} + \frac{n_2}{k_2} \right)  + 2\pi i n 
= \sum_{j\neq k } \ln \left( \frac{ 1 + \frac{i}{L(u_k- u_j)} }{ 1- \frac{i}{L(u_k -u_j)} }\right) . 
\end{equation}
Approximating the sum by an integral and expanding to $O(1/L)$ we obtain
\begin{equation}
\frac{1}{u} + 2\pi \left( n + \frac{n_1}{k_1} + \frac{n_2}{k_2} \right)
= 2 \pint dv \frac{\tilde \rho(v)}{ u-v},  
\end{equation}
where we have introduced $\tilde \rho$, the density of spins which 
satisfy the normalization given by 
\begin{equation}
 \int du \tilde \rho(u) = \frac{J}{L}. 
\end{equation}
The cyclicity constraint reduces to 
\begin{equation}
\int du  \frac{\tilde \rho(u)}{u}  =  -2\pi ( m + \frac{n_1}{k_1} ),
\end{equation}
and energy of the spin chain is then given by the expression
\begin{equation}
\label{spinen}
D =  \frac{ 2g^2 }{L} \int du \frac{\tilde \rho( u)}{u^2}  + O(g^4).
\end{equation}
Here again we have approximated the sum by an integral and also kept
the leading term in $g$

Now to compare these equations with that of the resolvent given in 
(\ref{comp1}),  (\ref{comp2}),  (\ref{comp3}) and (\ref{compjum})  we 
first need to identify the change of variables from the 
spectral parameter $u$ to $x$ in the sigma model. 
This was identified in  \cite{Beisert:2004hm} for the $SU(2)$ case and we use it for the Lens space. 
Let us redefine $u$ to be 
\begin{equation}
 u = x + \frac{g^{\prime 2}}{x}, \qquad g' = \frac{g}{L}. 
\end{equation}
Then the normalization of the density becomes 
\begin{equation}
\label{norm}
\int dx\left( 1- \frac{g^{\prime 2} }{x^2} \right) \tilde\rho( u(x) )  = \frac{J}{L}.
\end{equation}
To rewrite the Bethe equations in terms of the variable $x$ we need to
keep track of higher powers of $g$ in the LHS of the Bethe equations. 
One can show that 
\begin{equation}
\label{redefb}
L\ln\left(  \frac{x (Lu + \frac{i}{2} ) }{ x( Lu - \frac{i}{2} )}\right)  = 
\frac{1}{x} + \frac{g^{\prime 2}}{x^3} + \frac{g^{\prime 4}}{x^4} \sim 
\frac{x}{x^2- g^{\prime 2}}. 
\end{equation}
On substituting this in the equation (\ref{bethe2}) we obtain
\begin{eqnarray}
\label{bethe5}
\frac{x}{x^2 - g^{\prime 2} } + 2\pi \left( n + \frac{n_1}{k_1} + \frac{n_2}{k_2} \right)
&=& 2 \pint dy \left( 1- \frac{g^{\prime 2}}{y^2} \right) \frac{\tilde\rho( u(x)) }{
( x-y) ( 1- \frac{g^{\prime 2} }{xy}) }, \\ \nonumber
&=& 2\pint  \frac{\tilde\rho( u(y)) }{( x-y) }
- \frac{2g^{\prime 2}}{x} \int \frac{\tilde \rho( y) }{y^2}
+ O(g^{\prime 4} ).  
\end{eqnarray}
The cyclicity constraint in the variable $x$ reduces to 
\begin{equation}
\int \frac{\tilde\rho( u(x))}{x} = - 2\pi( m + \frac{n_2}{k_2} ),  
\end{equation}
where we have used (\ref{redefb}). 
Now to $O(g^2)$  the energy of the spin chain 
given in (\ref{spinen}) in terms of the variable $x$ can be written 
as 
\begin{equation}
\label{spinen1}
 D =  \frac{ 2g^2}{L} \int dx \frac{ \tilde \rho( u(x))}{x^2} + O(g^4).  
\end{equation}
Thus we have written the Bethe equations, the normalization condition of the 
density and the energy of the spin chain in terms of the parameter $x$. 

We will now show the correspondence of the Bethe equations and  the equations 
determining  the resolvent of the sigma model.  
Consider the difference of (\ref{comp1}) and (\ref{comp3}). We obtain
\begin{equation}
\label{norma}
\int dx \rho(x) ( 1 - \frac{1}{x^2} ) = \frac{4\pi}{\sqrt{\lambda}} J, 
\end{equation}
On performing the rescaling 
\begin{equation}
\label{rescale}
x\rightarrow \frac{4\pi L}{\sqrt\lambda} x, 
\end{equation}
the equation in (\ref{norma}) reduces to 
\begin{equation}
\int dx \rho(x) ( 1 -
\left( \frac{\sqrt\lambda}{4\pi L} \right)^2 \frac{1}{x^2} ) = 
\frac{J}{L}. 
\end{equation}
We see that  this equation is the same as the normalization condition 
(\ref{norm}) of the spin chain on the identification
\begin{equation}
\frac{\sqrt\lambda}{4\pi L } = g', \qquad \rho(x) = \tilde\rho(u(x) ).  
\end{equation}
We can now examine the  equation (\ref{compjum}).
We first rewrite (\ref{comp3}) using the rescaling in 
(\ref{rescale}) as 
\begin{equation}
\label{andim}
\frac{\Delta}{L} = 1 + 2 g^{\prime ^2} \int dx \frac{\rho(x) }{x^2}. 
\end{equation}
 Performing the 
rescaling in (\ref{rescale}) in (\ref{compjum})  and using the 
above equation  we obtain
\begin{equation}
\label{bethe4}
\pint d\xi \frac{\rho( \xi) }{ x-\xi} = \frac{x}{ x^2 - g^{\prime 2} }
+   \frac{2g^{\prime 2}}{x} \int \frac{\rho( y) }{ y^2} 
+ 2\pi\left( n + \frac{n_1}{k_1} + \frac{n_2}{k_2} \right)+ O( g^{\prime 4}) . 
\end{equation}
Now comparing (\ref{bethe4}) and ( \ref{bethe5}) we see that 
they agree to $O(g^{\prime 2)}$. 
Finally the expression for the energy of the spin chain is identical to the 
equation given in (\ref{spinen1}) reduces to the equation given in 
(\ref{andim}) on identifying the energy $D = \Delta -L$. 
Thus to $O(g^{\prime 4})$ we have shown that the
equations of the spin chain is identical to that satisfied by the 
resolvent. 
This completes our proof that to two loops,  the equations determining the 
resolvent of the Lens space sigma model are the same as that of the 
twisted long range spin chain.

\section{The BTZ black hole}

To set up notations and conventions we first review the construction 
of the BTZ black hole as a quotient of the $AdS_3$ hyperboloid given in 
\cite{Banados:1992gq}.  We will restrict our attention to the non-extremal 
black hole, the extremal case is discussed in appendix A. 
We first define the BTZ background as follows. Consider the hyperboloid defined by 
\begin{equation}
\label{btzconst}
 - u^2 - v^2 + x^2 + y^2 = - 1. 
\end{equation}
The BTZ black hole is constructed by the following parametrization of the 
hyperboloid

\vspace{.5cm}
\noindent
{\bf Region I } $r_+ < r$
\begin{eqnarray}
 u = \sqrt{A(r)} \cosh \tilde \phi (t, \phi), & & \qquad 
x= \sqrt{A(r)} \sinh \tilde\phi (t, \phi), \\ \nonumber
y = \sqrt{B(r)}\cosh \tilde t(t, \phi), & & \qquad 
v = \sqrt{B(r)} \sin \tilde t(t, \phi). 
\end{eqnarray}
\vspace{.5cm}
\noindent
{\bf Region II } $r_- < r< r_+$
\begin{eqnarray}
  u = \sqrt{A(r)} \cosh \tilde \phi (t, \phi), && \qquad
x= \sqrt{A(r)} \sinh \tilde\phi (t, \phi), \\ \nonumber
y = - \sqrt{- B(r)}\sinh \tilde t(t, \phi), && \qquad
v = - \sqrt{- B(r)} \cosh \tilde t(t, \phi). 
\end{eqnarray}
\vspace{.5cm}
\noindent
{\bf{Region II}} $0 < r< r_-$
\begin{eqnarray}
  u = \sqrt{-A(r)} \cosh \tilde \phi (t, \phi), && \qquad
x= \sqrt{-A(r)} \sinh \tilde\phi (t, \phi), \\ \nonumber
y = - \sqrt{- B(r)}\sinh \tilde t(t, \phi),  && \qquad 
v = - \sqrt{- B(r)} \cosh \tilde t(t, \phi), 
\end{eqnarray}
where 
\begin{equation}
\label{defabbtz}
 A(r) = \frac{ r^2 - r_-^2}{ r_+^2 - r_-^2}, \qquad 
B(r) = \frac{ r^2 - r_+^2}{ r_+^2 - r_-^2}, 
\end{equation}
and
\begin{equation}
 \tilde t = r_+ t - r_- \phi, \qquad \tilde\phi = - r_ - t + r_+\phi . 
\end{equation}
In the coordinates $r, t, \phi$ one obtains the metric
\begin{eqnarray}
ds^2 &=& - N^2 dt^2  + N^{-2} dr^2 + r^2 ( N^{\phi} dt  + d\phi )^2 , \\ \nonumber
N^2(r) &=& -M + {r^2} + \frac{ j^2}{4r^2}, \\ \nonumber
N^\phi(r) &=& -\frac{j}{2r^2}. 
\end{eqnarray}
The mass $M$ and the angular momentum $j$  of the BTZ black hole 
 are related to $r_+$ and $r_-$ by 
\begin{equation}
\label{rmj}
r_+^2 + r_-^2 = M , \qquad r_+r_- = \frac{j}{2}
\end{equation}
The metric is a solution of the equations of motion of the action
\begin{equation}
 I = \frac{1}{2\pi} \int d^3 x \sqrt{g} ( R + 2).
\end{equation}
Note that we have set the radius of $AdS_3$ to be unity. 
Further more note that in all the regions we have the following constraint. 
\begin{equation}
 A(r) - B(r) = 1. 
\end{equation}
The  BTZ solution as a quotient of the $AdS_3$ which is a
$SL(2, R)$ group manifold. This is seen as follows: 
 let us parametrize the $SL(2, R)$ 
group element as 
\begin{equation}
g = \left( \begin{array}{cc}
u+x & y+v \\
y-v & u-x
\end{array} \right), 
\end{equation}
with the constraint given in (\ref{btzconst}).  On rewriting the group element $g$ in terms of 
the variables $r, \tilde \phi, \tilde t$, the global coordinates of the BTZ metric we obtain 
the following parametrization of the $SL(2, R)$ group element in the various regions. 

\noindent
{\bf Region I:  $r>r_+$ }

\noindent
In the  region outside the horizon $g$  can be written as
\begin{equation}
\label{r1}
g = \left( 
\begin{array}{cc}
0 & -e^{\frac{1}{2} ( \tilde\phi + \tilde t ) } \\
e^{ - \frac{1}{2} ( \tilde \phi + \tilde t) } & 0 
\end{array}
\right)
\left( 
\begin{array}{cc}
\sqrt{A} &  - \sqrt{B} \\
-\sqrt{B} & \sqrt{A} \end{array}
\right)\left( 
\begin{array}{cc}
0  & e^{ \frac{1}{2} ( \tilde t - \tilde \phi) } \\
- e^{ - \frac{1}{2} ( \tilde t - \tilde \phi) } &0 
\end{array}
\right) .
\end{equation}

\noindent
{\bf Region II $ r_- < r< r_+$ }

\noindent
In the region between the inner and outer horizon  $g$ is given by 
\begin{equation}
\label{r2}
  g = \left( 
\begin{array}{cc}
0 & -e^{\frac{1}{2} ( \tilde\phi + \tilde t ) } \\
e^{ - \frac{1}{2} ( \tilde \phi + \tilde t) } & 0 
\end{array}
\right)
\left( 
\begin{array}{cc}
\sqrt{A} &  - \sqrt{-B} \\
\sqrt{-B} & \sqrt{A} \end{array}
\right)\left( 
\begin{array}{cc}
0  & e^{ \frac{1}{2} ( \tilde t - \tilde \phi) } \\
- e^{ - \frac{1}{2} ( \tilde t - \tilde \phi) } &0 
\end{array}
\right) .
\end{equation}

\noindent
{\bf Region III $0< r<r_-$ }

\noindent
Finally in the region inside the inner horizon $g$ is given by 
\begin{equation}
\label{r3}
  g = \left( 
\begin{array}{cc}
0 & -e^{\frac{1}{2} ( \tilde\phi + \tilde t ) } \\
e^{ - \frac{1}{2} ( \tilde \phi + \tilde t) } & 0 
\end{array}
\right)
\left( 
\begin{array}{cc}
-\sqrt{-A} &  - \sqrt{-B} \\
\sqrt{-B} & \sqrt{-A} \end{array}
\right)\left( 
\begin{array}{cc}
0  & e^{ \frac{1}{2} ( \tilde t - \tilde \phi) } \\
- e^{ - \frac{1}{2} ( \tilde t - \tilde \phi) } &0 
\end{array}
\right) .
\end{equation}
The quotienting of the group element arises because of the identification 
\begin{equation}
 \phi \sim \phi +2\pi.
\end{equation}
Under this identification and using (\ref{r1}), (\ref{r2}) and (\ref{r3}) it can be 
seen that in all regions,  the above identifications acts as 
\begin{equation}
\label{quobtz}
 g \sim \tilde A_{(1)} g A_{(1)},
\end{equation}
where 
\begin{equation}
\label{defak}
\tilde A_{(k)} = \left( 
\begin{array}{cc}
e^{( r_+ - r_-) \pi k } & 0 \\
0 & e^{- ( r_+ - r_-) \pi k } 
\end{array}
\right) 
\qquad
A_{(k)} = \left( 
\begin{array}{cc}
 e^{ ( r_- + r_+) \pi k } & 0  \\
0 & e^{- ( r_- + r_+) \pi k } 
\end{array}
\right). 
\end{equation}
We have now cast the quotienting of the $SL(2, R)$ group manifold which results in the 
BTZ black hole on similar lines to that of obtaining the Lens space from the $SU(2)$ group 
manifold discussed in the previous section. 

We will be interested in classical solutions only in the physical region I, 
that is the region outside the horizon. 
We will obtain classical solutions in 
the global coordinates $r, t, \phi$. Note that  in these coordinates 
any in falling geodesic approaches the horizon asymptotically.  
As we will see subsequently, this will be the situation
for the  solutions we  discuss in this paper. 
Using the parametrization of the $SL(2, R)$ group element 
given in (\ref{r1}), (\ref{r2}) and (\ref{r3}) 
the sigma model for  the  string propagating in BTZ times a $S^1$
is given by 
\begin{equation}
\label{sigact}
S = - \frac{\lambda}{2}  \int d^2 \sigma \left( \frac{1}{2}  {\rm Tr} 
( g^{-1} \partial_a g g^{-1} \partial^a g^{-1}
) + \partial_a Z \partial^a Z \right), 
\end{equation}
where $Z$ is the coordinate along the $S^1$  and  $\lambda$ is the coupling of the sigma model.
The equations of motion of the sigma model are given by
\begin{equation}
 \partial^\alpha (g^{-1} \partial g_a ) =0, \qquad   \partial^a \partial_a Z = 0. 
\end{equation}
As before let us define $j_a = g^{-1} \partial_a g$, then 
the Virasoro constraints are given by
\begin{equation}
\frac{1}{2} {\rm {Tr}} ( j_{\pm} j_\pm ) + \partial_\pm Z \partial_\pm Z =0. 
\end{equation}
The sum of the above two equations results in the Hamiltonian constraint
\begin{equation}
\label{ham}
\frac{1}{2} {\rm Tr} ( j_\tau j_\tau  + j_\sigma j_\sigma ) 
+ (\partial_\tau Z)^2 + (\partial_\sigma Z)^2 =0. 
\end{equation}
Note that $j_a$ is an $SL(2, R)$ current. Let us chose the generators of 
$SL(2,R)$ as 
\begin{equation}
t^1 = \sigma^1, \qquad t^2 = i \sigma^2, \qquad t^3=  \sigma^3. 
\end{equation}
Then we see that in  equation (\ref{ham}) the only negative 
definite quantity is the term
$- ( {\rm Tr}((  j_\tau^2) ^2 + ( j_\sigma^2) ^2) $, the remaining 
terms are positive definite. Thus for the 
Virasoro constraint to be satisfied we must have either
$j_\tau^2\neq 0$ or $j_\sigma^2\neq 0$.
This condition is important to note for the following reason:
unlike for the case of the $SU(2)$ as done in \cite{Kazakov:2004qf},  it is  not possible
for us to restrict the class of solutions for with only one component of the 
charge  say $\sigma^3$ is turned on.   We will elaborate on this in the next subsection. 
There are two consequences of the quotienting given in (\ref{quobtz}).
The first one is that  the sigma model admits solutions which have the 
twisted boundary condition 
 on the currents of the 
sigma model. 
\begin{equation}
j_{a }( \tau, \sigma+ 2\pi)  = A_{(k)}^{-1} j_{a}(\tau, \sigma) A_{(k)}. 
\end{equation}
Thus $k$ labels the winding number of the solution. 
The other is that in the quantum theory the charge corresponding to global  
shifts in $\phi$  quantized. 
We will not be dealing with the quantum theory so this condition 
is not relevant for the discussion in this paper. 
Now going through the same logic as in the case of the lens space we see that 
the sigma model on the BTZ space is integrable and the 
 the conserved charges can be extracted from the 
which monodromy matrix given by 
\begin{equation}
\label{fullmono}
\hat \Omega_k(x)  = A_{(k)} 
P \exp \left[ 
- \int_0^{2\pi} d\sigma \frac{1}{2} \left( \frac{j_+}{1-x} - \frac{j_-}{ 1+x} \right) \right]. 
\end{equation}
Since the monodromy matrix is an $SL(2, R)$ group element its
 eigen values are of the form 
$\{ \exp( ip (x) ) ,  \exp( -ip(x) ) \}$.   $p(x)$ is called 
the quasi-momentum and it characterizes the monodromy matrix. 
We also have the relation
\begin{equation}
 {\rm Tr}( \Omega (x) ) = 2 \cos p(x).  
\end{equation}
We will see that corresponding to each classical solution there exists a density  which 
characterizes this solution in the  spectral plane. 
To proceed further we will define the  global charges 
which will play a special role in our analysis. 
The energy which corresponds to the global translations in time $t$ is given by 
\begin{equation}
 E =  \frac{\lambda}{4}  \left[ \int_0^{2\pi} d\sigma \left(  ( r_+ - r_-) {\rm Tr} ( \partial_0 g g^{-1} \sigma^3 ) - 
 ( r_+  + r_-) {\rm Tr} ( g^{-1}\partial_0 g \sigma^3)  \right) \right] . 
\end{equation}
The spin which corresponds to the global translations in $\phi$ is given by 
\begin{equation}
 S =  \frac{\lambda }{4} \left[ \int_0^{2\pi} d\sigma \left(  ( r_+ - r_-) {\rm Tr} ( \partial_0 g g^{-1} \sigma^3 ) +
 ( r_+ +  r_-) {\rm Tr} ( g^{-1}\partial_0 g \sigma^3)  \right) \right]. 
\end{equation}
Thus the following combinations of charges  have a simple relation in terms of the 
right and left currents. 
\begin{eqnarray}
\label{btzc1}
 E+ S =\frac{\lambda}{2}  ( r_+ -  r_-) \int_0^{2\pi} 
 d\sigma {\rm Tr} ( \partial_0 g g^{-1} \sigma^3 ), \\ 
 \label{btzc2}
E - S =  -\frac{\lambda}{2}  ( r_+ + r_-) \int_0^{2\pi} 
 d\sigma {\rm Tr} ( g^{-1}\partial_0 g \sigma^3). 
\end{eqnarray}
Global translations in the coordinate $Z$ leads to the charge  $\hat J$ which is 
given by 
\begin{equation}
 \hat J = \lambda \int d^2 \sigma \partial_0 Z, 
\end{equation}
We will choose a gauge in which 
\begin{equation}
 Z = \frac{\hat J}{2\pi \lambda} \tau + \hat m \sigma, 
\end{equation}
where $\hat m$ is the winding number, this is the same gauge chosen 
in \cite{Kazakov:2004nh}.  The Virasoro constraints then reduce to 
\begin{equation}
\label{virbtz}
 {\rm Tr} ( j_\pm^2) = 2 \left( \frac{\hat J}{2\pi \lambda} \pm \hat m \right). 
\end{equation}

Before we begin our analysis of the quasi-momentum we mention that 
the sigma model on $AdS_3\times S^1$ was studied earlier with a 
parametrization of the global $AdS_3$ coordinates by \cite{Kazakov:2004nh}. For this parametrization 
the right and left charges corresponding to the $t^2 = i\sigma^2$ generator,
the compact direction  played an important role \footnote{See equation (3.13) of the paper.}.

\subsection{Properties of the quasi-momentum}

In this subsection we discuss the  properties of the quasi-momentum in 
the spectral plane. This will enable us to demonstrate that every classical 
solution of the sigma model is characterized by a density in the complex $x$ plane. 
The properties of the quasi-momentum obtained from general considerations
will be confirmed in the next subsection by explicitly solving for the quasi-momentum 
for two class of solutions. 
To discuss the properties of the quasi-momentum we need to 
be careful of two sectors of the theory. The sector with winding zero $k=0$ and the 
sector with $k\neq 0$. We start with the zero winding sector. 

\vspace{.5cm}
\noindent
{\bf{Sector $k=0$}} 

Just as in the case of Lens space we will see that the quasi-momentum 
has a pole at $x=\pm1$. 
Let us first examine the case $x\rightarrow 1$. In this limit, we see that the 
monodromy matrix becomes
\begin{equation}
 \hat\Omega (x) \rightarrow  P \left[\exp\left(  - \int_0^{2\pi} d\sigma \frac{1}{2} \frac{j_+}{1-x} \right)
\right], 
\qquad x\rightarrow 1, 
\end{equation}
For this path ordered exponential to make sense,   the integrability  condition of the 
first order equation satisfied by $\hat \Omega$ implies that 
\begin{equation}
 \partial_- j_+ =0.
\end{equation}
Then from the equation of motion (\ref{eom1}) and the equation (\ref{flat1})  we obtain 
\begin{equation}
 \partial_+ j_- =0, \qquad [j_+, j_-] =0. 
\end{equation}
A simple solution to these equations is that  $j_+, j_-$ approach a constant as $x\rightarrow 1$ and 
we can choose $j_\pm$ to point along the same direction in the group space. 
To satisfy the Virasoro constaint in (\ref{virbtz}) is is convenient to choose
 choose $j_+$ to be along $i \sigma^2$.  Thus we obtain
\begin{equation}
j_+ =  i \left(\frac{1}{2\pi \lambda} \hat  J+\hat m\right ) \sigma^2, \qquad  x\rightarrow +1
\end{equation}
where we have chosen the positive square root.  
Note that this form for $j_+$ also satisfies the condition $\partial_- j_+ =0$. 
This means that the monodromy matrix reduces to 
\begin{equation}
\Omega \sim \exp \left(- i \frac{ \frac{\hat J}{2\pi \lambda} +\hat m  }{1-x}  \pi \sigma^2  \right), 
\qquad x\rightarrow 1.
\end{equation}
Thus we conclude that the quasi-momentum at $x\rightarrow 1$ is given by
\begin{equation}
\label{kz1}
p \sim \pi \frac{ \frac{\hat J}{2\pi \lambda} +\hat m }{ x-1}.
\end{equation}
A similar analysis as $x\rightarrow -1$ gives the following behaviour 
of the quasi-momentum 
\begin{equation}
\label{kz2}
p \sim \pi \frac{\frac{\hat J }{2\pi \lambda} -\hat m} { x+1}.
\end{equation}
Thus we have shown that the quasi-momentum has poles at $x=\pm 1$ with residues determined
by the charge $\hat J$ and the winding number $\hat m$. 
Now let us examine the behaviour of the monodromy matrix as $x\rightarrow \infty$.  
At $x\rightarrow \infty$ the expansion of $J_\sigma$ is given in (\ref{infexp}). 
Substituting  this expansion in the monodromy matrix we obtain
\begin{eqnarray}
\label{kzinf}
 2\cos p(x) &=&  2 + \frac{1}{2x^2} 
\int_0^{2\pi} d\sigma d\sigma' {\rm Tr} ( j_\tau(\sigma) j_\tau(\sigma')  ) + \cdots, 
\\ \nonumber 
&=& 2 + \frac{1}{4x^2} \left((  Q_R^1)^2 -  ( Q_R^2)^2 +  ( Q_R^3)^2 \right)  + \cdots, 
\end{eqnarray}
where we have defined
\begin{equation}
 Q_R^{i} = \int_0^{2\pi}  d\sigma {\rm Tr} \left( g^{-1} \partial_\tau g t^i\right).
\end{equation}
Let us denote the invariant
\begin{equation}
 Q_R^2 = (  Q_R^1)^2 -  ( Q_R^2)^2 +  ( Q_R^3)^2.
\end{equation}
Then from the equation (\ref{kzinf}) we have 
\begin{equation}
\label{kzinf1}
 p(x) \rightarrow \frac{i }{2x} \sqrt{ Q_R^2}  , \qquad  x\rightarrow \infty. 
\end{equation}
At this point it is worthwhile to point out a  difference in the analysis of 
\cite{Kazakov:2004qf} and \cite{Kazakov:2004nh}. 
There they assume that among the three global charges 
only one of them 
 contribute \footnote{See equations (4.38) and (3.27) of the respective papers.}, 
here we retain the dependence 
on all the three charges. As we will see subsequently that 
geodesics of  the BTZ background in general carry all the three global charges and it 
is necessary to retain their dependence to discuss all solutions.  
Now we can examine the behaviour as $x\rightarrow 0$. 
Using the relation in (\ref{zeroexp}) we can write the monodromy matrix as
\begin{equation}
 \Omega(x) = g^{-1}(2\pi) P \exp
 \left( - x\int_0^{2\pi} \partial_\tau g g^{-1} \right) g(0).
\end{equation}
We then use the fact that $g(2\pi) = g(0)$ in the untwisted sector to obtain 
\begin{equation}
 2\cos p(x) = 2  + \frac{x^2}{4}  \left((  Q_L^1)^2 -  ( Q_L^2)^2 +  ( Q_L^3)^2 \right), 
\end{equation}
where 
\begin{equation}
 Q_L^i = \int_0^{2\pi}  d\sigma {\rm Tr} ( t^{i}   \partial_{\tau} g g^{-1}  ) .
\end{equation}
Thus the behaviour of the quasi-momentum as $x\rightarrow 0$ is given by 
\begin{equation}
\label{kzero}
 p(x) \rightarrow  2\pi  m  + i \frac{x}{2} \sqrt{ Q_L^2}, x\rightarrow 0, 
\end{equation}
where 
\begin{equation}
 Q_L^2 = (  Q_L^1)^2 -  ( Q_L^2)^2 +  ( Q_L^3)^2. 
\end{equation}
Here again we have retained the dependence on all the three components of the 
global charges. 
Across branch cuts the quasi-momentum satisfies the equation
\begin{equation}
\label{kzjump}
 p(x+ i \epsilon) + p( x-i\epsilon)  = 2\pi n_l, 
\end{equation}
where $n_l$ is an integer for the $l$th cut. 
The reason for this is the same as in the case of the $SU(2)$, across branch 
cut the quasi-momentum 
takes the two  possible different eigen values of the monodromy matrix. 
The condition in (\ref{kzjump}) just arises from the 
unimodularity of the monodromy matrix. 

Let us now recast these properties of the quasi-momentum in terms of the resolvent. 
We can define the resolvent as
\begin{equation}
 G(x) = p(x) - \pi \frac{\frac{\hat J }{2\pi \lambda} +\hat m}{x-1} - 
\pi \frac{ \frac{\hat J}{2\pi \lambda} -\hat m}{x+1}.
\end{equation}
This ensures that the resolvent is a function in the complex plane without 
poles. Thus using standard complex analysis one can write the 
resolvent as 
\begin{equation}
G(x) = \int d\xi \frac{\rho(x) }{ x - \xi},
\end{equation}
where the integral is along the cuts. 
In fact $\rho$ is given by
\begin{equation}
\rho(x) = \frac{1}{2\pi i } \left( G( x+ i\epsilon) - G( x- i\epsilon) \right).
\end{equation}
Now from the behaviour of the quasi-momentum at $x\rightarrow \infty$ we obtain
\begin{equation}
\label{rho1}
 \int d\xi \rho(\xi) = -\frac{\hat J}{\lambda}  + \frac{i}{2} \sqrt{ Q_R^2}.
\end{equation}
>From the behaviour of the quasi-momentum at $x\rightarrow 0$ we obtain 
the following conditions on the density
\begin{eqnarray}
\label{rho2}
 \frac{1}{2\pi i} \oint dx \frac{G(x)}{x}  &=& - \int d\xi \frac{\rho(\xi)}{\xi} = 2\pi(  m + \hat m) , 
\\  
\label{rho3}
 \frac{1}{2\pi i } \oint dx \frac{G(x)}{x^2}  &=&  - \int d\xi \frac{\rho(\xi)}{\xi^2}
= \frac{J}{\lambda} + \frac{i}{2} \sqrt{ Q_L^2}.
\end{eqnarray}
Finally the condition in (\ref{kzjump}) for the resolvent across branch cuts gives rise to 
\begin{eqnarray}
\label{rho4}
G(x+ i \epsilon) + G(x- i\epsilon) &=& 2\pint d\xi \frac{\rho(\xi)}{x-\xi}, \\ \nonumber
&=& - \frac{ 2\pi ( \frac{\hat J }{2\pi \lambda} +\hat m)}{ x -1} - 
\frac{2\pi (\frac{\hat J }{2\pi \lambda} -\hat m)}{ x+1} + 2\pi n_l .
\end{eqnarray}
Equations (\ref{rho1}), (\ref{rho2}), (\ref{rho3}) and (\ref{rho3})  
show that give a classical solution in the $k=0$ sector, it 
determines a density in the spectral plane $x$.

\vspace{.5cm}
\noindent
{\bf{Sector $k \neq 0$} } 

Let us now repeat the analysis for the twisted sectors. 
The behaviour of the monodromy matrix as $x\rightarrow \pm1$ is same as in the case of $k=0$. 
From the earlier analysis it is easy to see that the presence of the matrix 
$A_k$ for the $k\neq 0$ does not affect the singular behaviour of the 
quasi-momentum as $x\rightarrow \pm 1$. Thus we have
\begin{equation}
\label{k1}
p \rightarrow  \pi \frac{  \frac{\hat J }{2\pi \lambda} \pm \hat m }{ x\mp 1}, \qquad x\rightarrow\pm 1
\end{equation}
As $x\rightarrow\infty$ we can use the expansion given in (\ref{infexp}) to obtain the 
following expression for the quasi-momentum  
\begin{eqnarray}
2 \cos p(x) &=& {\rm Tr} \left( A_{(k)} P \exp\int_0^{2\pi} ( d\sigma \frac{j_0}{x} + \frac{j_1}{x^2} \cdots)  \right)
\\ \nonumber
 &=& {\rm Tr} \left[(  \cosh \pi k ( r_+ + r_-) + \sigma^3 \sinh \pi k( r_+ + r_-) ) \right. 
\\ \nonumber
& \times& \left. \left ( 1+ \frac{t^i}{x} \int_0^{2\pi} d\sigma j_0^i  + 
\frac{t^i t^j}{x^2} \int_0^{2\pi}d\sigma \int_0^{\sigma} d\sigma'
j_0^i(\sigma)  j_0^j (\sigma') 
+ \frac{t^i}{x^2} \int_0^{2\pi} d\sigma j_1^i(\sigma)  \right) \right]
 \\ \nonumber
&=& 2 \cosh\pi k ( r_+ + r_-)   +  \frac{1}{x} \sinh\pi k ( r_+ + r_-) 
 {\rm Tr}( \int_0^{2\pi} d\sigma j_0 \sigma ^3)+\frac{{\cal C}} {x^2} + \cdots ),  \\ \nonumber
& =& 2 \cosh\pi k ( r_+ +r_-)   -  \frac{1}{x} \sinh\pi k ( r_+ + r_-) 
\frac{2(E-S)}{\lambda(r_+ + r_-)}
+ \frac{{\cal C}}{x^2} + \cdots.  
\end{eqnarray}
In the last line we have used the definition of the global charge given in 
(\ref{btzc2}).  Here we have expanded the currents as
\begin{equation}
 g^{-1} \partial_a g = t^i j_a^i .
\end{equation}
The first non-local charge ${\cal C}$ is given by 
\begin{eqnarray}
\label{nonlocal1}
{\cal C} &=& 2 \cosh \pi( r_+ + r_-) \int_0^{2\pi}d\sigma \int_0^{\sigma} d\sigma'
j_0^i(\sigma) j_0^i(\sigma') \\ \nonumber & & 
- 2 \sinh\pi(r_+ + r_-) \left( \int_0^{2\pi}d\sigma \int_0^{\sigma} d\sigma'
j_0^{[1, }(\sigma ) j_0^{2]}(\sigma') - 
\int_0^{2\pi} d\sigma j_1^3(\sigma) \right), 
\label{order}
\end{eqnarray} 
where  $j_0^i j_0^i = j_0^1 j_0^1 - j_0^2 j_0^2  + j_0^3 j_0^3$.
Thus the leading behaviour of the quasi-momentum is determined by the 
global charges $E, S$ and is given by 
\begin{equation}
\label{beinf}
p( x) \sim  i \pi k ( r_+ + r_-) - i \frac{1}{x} \frac{E-S}{\lambda( r_+ + r_-)}
\end{equation}
At this point it is relevant to point out the difference in behaviour for the
$k=0$ case.  From (\ref{kzinf1}) one see that the all the three components
of the global charges determine the $O(1/x)$ term unlike the case above. 
Therefore settting $k=0$ in the expression (\ref{beinf}) does not reduce to 
(\ref{kzinf1}) unless we restrict  to the situation in which charges 
corresponding to the $\sigma^3$ direction is turned on and others are set to
zero. 
Note that the two leading terms in quasi-momentum 
is purely imaginary as $x\rightarrow \infty$. 
Now let us examine the limit $x\rightarrow 0$, again using 
(\ref{zeroexp}) we get
\begin{equation}
\Omega(x) = A_{(k)} g^{-1} (2\pi) P \exp
 \left( - \int_0^{2\pi} d\sigma (x \partial_\tau g g^{-1} + x^2 \partial_\sigma g g^{-1} ) \right) g(0), 
\end{equation}
where we have retained the $O(x^2)$ term also. 
Taking the trace  and using the fact
\begin{equation}
g(2\pi) = \tilde A_{(k)} g(0) A_{(k)}, 
\end{equation}
we obtain
\begin{eqnarray}
2 \cos p(x) &=&
{\rm Tr}( \tilde A_{(k)}^{-1} \exp 
 \left( - \int_0^{2\pi} d\sigma (x \partial_\tau g g^{-1} + x^2 \partial_\sigma g g^{-1} ) \right), 
 \\ \nonumber
&=& {\rm Tr}
[  ( \cosh \pi k ( r_+ - r_-) - \sigma^3 \sinh \pi k ( r_+ - r_-) )  \\ \nonumber
& \times&  \left.  \left(
1 - x t^i \int_0^{2\pi} d\sigma l^i_0 + x^2 t^it^j \int_0^{2\pi} \int_0^{\sigma'} d\sigma'
l_0^i(\sigma) l_0^j(\sigma') - x^2 t^i \int_0^{2\pi} d\sigma l_1^i \right) \right], \\ \nonumber
&=&2 \cosh \pi k ( r_+ - r_-) + 2x \sinh \pi k ( r_+ - r_-) \frac{E + S}{\lambda( r_+ - r_-)}+
x^2{\cal Q}+ \cdots,  \\ \nonumber
\end{eqnarray}
where we have used (\ref{btzc1}) to obtain the last line.  Here we have expanded the currents as
\begin{equation}
 \partial_a g g^{-1} = t^i l_a^i,
\end{equation}
For completeness we write down the 
non-local charge obtained at $O(x^2)$ which is given by  
\begin{eqnarray}
\label{nonlocal2}
{\cal Q} &=& 2\cosh \pi k ( r_+ + r_-)\int^{2\pi}_0 d\sigma 
\int^{\sigma}_0 d\sigma' l_0^i(\sigma)l_0^i(\sigma ') \\ \nonumber
& & \left. + 2 \sinh \pi k ( r_+ + r_-) \left( \int^{2\pi}_0 d\sigma \int^{\sigma}_0 d \sigma 'l_0^{[1,}(\sigma)l_0^{2]}(\sigma ') + \int^{2\pi}_0 d\sigma l_1^3 \right) \right.
\end{eqnarray}
Thus the leading  behaviour of the 
quasi-momentum at $x\rightarrow 0$ is given by  
\begin{equation}
\label{pzero}
p ( x) \sim  2\pi  m + i \pi k ( r_+ - r_-) + i x   \frac{E + S}{\lambda( r_+ - r_-) }.
\end{equation}
Again we see that the quasi-momentum is purely imaginary as
$x\rightarrow 0$.  Furthermore unlike the case of $k=0$, see equation (\ref{kzero}),  the leading behaviour 
depends on only on the third component of the global charge.
The unimodularity of the monodromy matrix imposes the following condition 
on the quasi-momentum across branch cuts. 
\begin{equation}
\label{jumcon}
p(x + i\epsilon) + p( x- i \epsilon) = 2\pi n_l, 
\end{equation}
where $n_l$ refers to an integer corresponding to the $l$th branch cut. 

Let us now use all these information and obtain the conditions on the 
resolvent. We define the resolvent as 
\begin{equation}
G(x) = p(x)  - \pi \frac{\frac{\hat J}{2\pi \lambda} +\hat m}{x-1} - 
\pi \frac{ \frac{\hat J}{2\pi \lambda} -\hat m}{x+1} - i \pi k ( r_+ + r_-). 
\end{equation}
This ensures that the resolvent is a function on the complex plane without 
any poles and falls off as $1/x$ for large values of the spectral parameter. 
By standard complex analysis we can write the resolvent in terms of a density function by 
\begin{equation}
G(x) = \int d\xi \frac{\rho(x) }{ x - \xi}, 
\end{equation}
here the integral is along the various cuts. 
Then from the behaviour of the quasi-momentum at $x\rightarrow\infty$
given in (\ref{beinf})  we obtain
\begin{equation}
\label{den1}
\int d\xi \rho(\xi) = -\frac{\hat J}{\lambda}   -  i \frac{E-S}{ \lambda( r_+ + r_-) }. 
\end{equation}
From the behaviour of the quasi-momentum at 
$x\rightarrow 0$ given in (\ref{pzero}) we obtain the following two equations
\begin{eqnarray}
\label{den2}
 \frac{1}{2\pi i } \oint dx \frac{ G(x)}{x} 
&=& - \int  d\xi \frac{\rho(x)}{\xi} = 
 2\pi ( \hat m + m  - i  k r_-),   \\ 
\label{den3}
 \frac{1}{2\pi i } \oint dx \frac{ G(x) }{x^2} 
&=& - \int  d\xi \frac{\rho(\xi)}{ \xi^2} =  \frac{ \hat J }{\lambda}   + i \frac{E+ S}{\lambda( r_+ - r_-)}.
\end{eqnarray}
Finally we have to use the unimodularity of the monodromy matrix  in (\ref{jumcon}) to get  
\begin{eqnarray}
\label{den4}
 G(x+ i \epsilon) + G(x- i\epsilon) &=& 2\pint d\xi \frac{\rho(\xi)}{x-\xi}, \\ \nonumber
&=& - \frac{ 2\pi ( \frac{\hat J }{2\pi \lambda} +\hat m)}{ x -1} - 
\frac{2\pi (\frac{\hat J }{2\pi \lambda} -\hat m)}{ x+1} + 2\pi n_l - 2\pi i k( r_+ + r_-). 
\end{eqnarray}
Thus we have completed the proof that 
a classical solution of of the sigma model in the BTZ background corresponds to a 
density function 
in the complex plane. The density is determined by solving the above set of 
integral equations.

\subsection{Classical Solutions and quasi-momentum}

In this subsection we study two  examples of classical solutions of the 
BTZ sigma model and evaluate the quasi-momentum explicitly. 
We will show that the behaviour of the quasi-momentum agrees 
with the general discussion in the previous subsection for these examples. 

To obtain classical solutions it is convenient to write the BTZ sigma model 
in terms of the coordinates of the $AdS_3$ hyperboloid $u, v, x, y$ and impose the 
constraint given in (\ref{btzconst}). 
Then the sigma model action is given by 
\begin{eqnarray}
\label{coord-act}
 S &=& - \frac{\lambda}{2} \int d\tau d\sigma 
\left[ - \partial_a u \partial^a u 
- \partial_a v \partial^a v 
 + \partial_a x \partial^a x 
+ \partial_a y \partial^a y \right. \\ \nonumber
& & \left. - \Lambda (  - u^2 - v^2 + x^2 + y^2 +1) +  \partial_a Z\partial^a Z \right], 
\end{eqnarray}
where $\Lambda$ is the Lagrange multiplier. 
The equations of motion are given by 
\begin{eqnarray}
& & -  \partial^a \partial_a u =  \Lambda u, \qquad
- \partial^a\partial _a v = \Lambda   v, \\ \nonumber
& & \partial^a\partial_a x = - \Lambda x, \qquad
\partial^a\partial_a y = - \Lambda y, \qquad   \partial^a\partial_a Z = 0. 
\end{eqnarray}
The Virasoro constraints in these variables are given by 
\begin{equation}
  - \partial_\pm u \partial_\pm u - \partial_\pm v \partial_\pm v 
+ \partial_\pm x  \partial_\pm x  + \partial_\pm y \partial_\pm y  
+ \partial_\pm Z\partial_\pm Z =0.
\end{equation}
Since we are interested in closed string solutions, all solutions are subject to the periodicity conditions
\begin{equation}
\label{period}
 r(\tau, \sigma +2\pi) = r(\tau, \sigma), \qquad
t(\tau, \sigma + 2\pi) = t(\tau, \sigma), \qquad
\phi(\tau, \sigma + 2\pi) = \phi( \tau, \sigma) + 2\pi k 
\end{equation}
where $k$ is the winding number. 
With these preliminaries we are ready to discuss the classical solutions. 
Our solutions fall into two classes, the geodesics and the winding strings.

\vspace{.5cm}
\noindent
{\bf Geodesics}

We first consider solutions which are independent of the world sheet 
variable $\sigma$. 
We will show that these solutions correspond to geodesics in the BTZ background. 
For this we consider the following ansatz.  
\begin{eqnarray}
 u+x = a(\tau) \exp( f(\tau) ), \qquad u-x = a(\tau) \exp( -f(\tau)) , \\ \nonumber
y +v =  b(\tau) \exp( g(\tau)), \qquad y -v = b(\tau) \exp( -g(\tau)).
\end{eqnarray}
Substituting this ansatz in the action given in (\ref{coord-act}) we obtain the following 
Lagrangian
\begin{equation}
 \frac{1}{2} \left( ( \dot a)^2 - ( \dot b)^2 - ( \dot f)^2  a^2 + ( \dot g )^2 b^2 -\Lambda ( -a^2 + b^2 + 1)  
\right).
\end{equation}
We can  eliminate the constraint by the following substitution
\begin{equation}
 a(\tau) = \cosh \gamma(\tau), \qquad b(\tau) = \sinh \gamma(\tau).
\end{equation}
The equations of motion then can be obtained from the Lagrangian 
\begin{equation} 
 L = \frac{1}{2} \left( - (\dot \gamma)^2 - (\dot f)^2  \cosh^2 \gamma +  ( \dot g)^2  \sinh^2\gamma \right)
\end{equation}
From here we see that  there are two constants of motion given by 
\begin{equation}
 \dot f \cosh^2 \gamma = c_1, \qquad \dot g \sinh^2 \gamma = c_2. 
\end{equation}
These constants are related to the charges $E$ and $S$ as follows
\begin{eqnarray}
\label{eands}
 E- S =  -( r_+ + r_-) 2\pi \lambda  ( c_1 + c_2), \\ \nonumber
E+ S = ( r_+ - r_-) 2\pi \lambda ( c_1 - c_2).
\end{eqnarray}
Substituting these constants of motion we obtain  the following equation of motion for $\gamma$
\begin{equation}
\ddot \gamma = - \frac{\sinh\gamma}{\cosh ^3 \gamma} c_1^2 + \frac{\cosh\gamma}{\sinh^3 \gamma} c_2^2.
\end{equation}
The Virasoro constraints can be used to integrate this equation to quadratures. This is given by
\begin{equation}
\label{integral1}
 ( \dot \gamma)^2 + \frac{c_1^2}{\cosh^2 \gamma} - \frac{c_2^2}{\sinh^2\gamma} + 
( \frac{\hat J}{2\pi\lambda}  \pm \hat m)^2 = 0.
\end{equation}
The two equations imply either $\hat J=0$ or $\hat m =0$. Let us assume $\hat m =0$ therefore we obtain
\begin{equation}
\label{potential}
 ( \dot \gamma)^2 + \frac{c_1^2}{\cosh^2 \gamma} - \frac{c_2^2}{\sinh^2\gamma} + 
( \frac{\hat J}{2\pi\lambda} )^2 = 0.
\end{equation}
It is clear from the above equation that there are always initial conditions
which allow for real  solutions. 
We will assume the initial condition that at $\tau =0$   we have  $\gamma_0$ satisfying 
\begin{equation}
  \frac{c_1^2}{\cosh^2 \gamma_0} - \frac{c_2^2}{\sinh^2\gamma_0} + 
( \frac{\hat J}{2\pi\lambda} )^2 = 0, 
\end{equation}
and $\dot\gamma =0$.  We can also assume that $\tau=0$, the coordinates $t=0, \phi=0$. 
From the structure of the potential in (\ref{potential}) we see that the geodesic always falls into the 
horizon.  One can explicitly integrate the equation for $\gamma$.
In Appendix B we show that the equation (\ref{potential}) corresponds to the 
equation of a geodesic in the BTZ background.

Let us now evaluate the monodromy matrix for this solution to verify the general properties discussed in 
the earlier section.  Note that this solution is in the $k=0$ sector, since there is no winding. 
Evaluating the current $g^{-1}\partial_\pm g $ on this solution we obtain
\begin{equation}
g^{-1}\partial_\pm g  = 
\left(\begin{array}{cc}
c_1 + c_2 &  
( \dot\gamma + \frac{\sinh\gamma}{\cosh\gamma} c_1 + \frac{\cosh\gamma}{\sinh\gamma} c_2 )
e^{ g-f} \\ 
( \dot\gamma - \frac{\sinh\gamma}{\cosh\gamma} c_1 - \frac{\cosh\gamma}{\sinh\gamma} c_2 )
e^{ f-g} & - ( c_1 +c_2). 
\end{array}
\right).
\end{equation}
Since the currents do not depend on the world sheet coordinate $\sigma$ it is 
easy to perform the path ordered integral for the monodromy matrix. We obtain 
\begin{eqnarray}
\Omega &=& \exp \left[ - \frac{2\pi x}{1-x^2}   g^{-1} \dot g  \right],
\\ \nonumber
&=& \cos \theta + i \hat \theta \cdot \sigma \sin \theta, 
\end{eqnarray}
where the 3-vector $\theta$ is given by 
\begin{eqnarray}
 \vec\theta  = 
\left( \begin{array}{c}
        - i \left[ \dot\gamma \cosh( g-f) + 
\left(   \frac{\sinh\gamma}{\cosh\gamma} c_1 + \frac{\cosh\gamma}{\sinh\gamma} c_2 \right) 
\sinh( g-f) \right] \\
\dot\gamma \sinh( g-f) + 
\left(   \frac{\sinh\gamma}{\cosh\gamma} c_1 + \frac{\cosh\gamma}{\sinh\gamma} c_2 \right) 
\cosh( g-f)  \\
-i( c_1 + c_2) 
       \end{array}
\right).
\end{eqnarray}
Let us evaluate the (modulus)$^2$ of this three vector,  it is given by
\begin{eqnarray}
\theta^2 &=& (\frac{2\pi x}{ x^2-1} ) ^2 \left( 
-( \dot \gamma)^2 - \frac{c_1^2}{\cosh^2\gamma} + \frac{c_2^2}{\sinh^2}  \right), \\ \nonumber
& =& 
(\frac{2\pi x}{ x^2-1} ) ^2 \frac{\hat J^2}{4\pi^2 \lambda^2}. 
 \end{eqnarray}
 In the second line we have used the Virasoro constraint given in (\ref{potential}) 
>From this property it is easy 
to evaluate the   eigen values of this monodromy matrix.
They are given by  are given by $\exp( \pm i \theta)$. This results in the 
following values of quasi-momentum on this solution
\begin{equation}
\label{solpgeo}
p = \frac{\hat J}{\lambda} \frac{x}{ x^2-1}, 
\end{equation}
where we  have taken the positive  square root. 
Thus the behaviour of the quasi-momentum for $x\rightarrow \pm 1$ is 
given by 
\begin{eqnarray}
p \rightarrow  \frac{ \hat J}{2\lambda} \frac{1}{x\mp1}, \qquad x\rightarrow \pm1, 
\end{eqnarray}
One can see that the the  solution for the quasi-momentum for 
geodesics satisfy the equations (\ref{kz1}) and (\ref{kz2}). 
We also see from the formula for the resolvent in (\ref{solpgeo}) we see that $G(x)=0$ for this 
case, this implies the density vanishes. 
The charges  $Q_R$ and $Q_L$ for the geodesic solutions are  given by
\begin{eqnarray}
 Q_R^1 &=& 4\pi \left( \dot\gamma \cosh(g-f) +
 \left(   \frac{\sinh\gamma}{\cosh\gamma} c_1 + \frac{\cosh\gamma}{\sinh\gamma} c_2 \right) 
\sinh( g-f) \right) ,\\ \nonumber
Q_R^2 &=& 4\pi \left( \dot\gamma \sinh( g-f) + 
\left(   \frac{\sinh\gamma}{\cosh\gamma} c_1 + \frac{\cosh\gamma}{\sinh\gamma} c_2 \right) 
\cosh( g-f)\right), \\ \nonumber
Q_R^3 &=& 4\pi ( c_1 + c_2), 
\end{eqnarray}
Similarly the charges left charges are given by
\begin{eqnarray}
 Q_L^1 &=& 4\pi \left(  \dot\gamma \cosh(g+f) -
 \left(   \frac{\sinh\gamma}{\cosh\gamma} c_1 + \frac{\cosh\gamma}{\sinh\gamma} c_2 \right) 
\sinh( g-f) \right),\\ \nonumber
Q_L^2 &=& 4\pi \left(\dot\gamma \sinh( g+f) -
\left(   \frac{\sinh\gamma}{\cosh\gamma} c_1 + \frac{\cosh\gamma}{\sinh\gamma} c_2 \right) 
\cosh( g+f)\right) , \\ \nonumber
Q_L^3 &=&  4\pi  \left( c_1 - c_2\right) , 
\end{eqnarray}
One can evaluate the norms of these charges and show 
\begin{equation}
 Q_R^2  = Q_L^2 = - \frac{4 \hat J^2}{\lambda^2}. 
\end{equation}
This ensures that the equations (\ref{rho1}) and (\ref{rho2}) are satisfied for zero density. 
Since the form for the quasi-momentum given in (\ref{solpgeo}) does not have a branch cut, the 
equation (\ref{kzjump}) does not arise. 
One can do a similar analysis for the situation in which $\hat J=0$ and $\hat m\neq 0$ and show
that all the equations obtained for the quasi-momentum from general considerations
hold. At this point we mention that these solutions in general have all the components of the 
charge turned on unlike the solutions considered in \cite{Kazakov:2004qf} and 
\cite{Kazakov:2004nh} in which classical 
solutions carried only one component of the charge.  Note that  geodesics in the 
BTZ background   can be thought of as a vacuum since the corresponding density in 
the spectral plane  vanishes.

\vspace{.5cm}
\noindent
{\bf Winding strings}

In this section we discuss classical strings 
which wind around the $\phi$ direction and the corresponding
quasi-momentum. For this we start with the ansatz
\begin{eqnarray}
 u+x = a(\tau) e^{( f(\tau) + \nu_1\sigma) }, \qquad u -x = a(\tau) e^{-( f(\tau) + \nu_1\sigma) }, 
\\ \nonumber
y+v = b(\tau) e^{( g(\tau) + \nu_2 \sigma)}, 
\qquad
y-v = b(\tau) e^{ - ( g(\tau) + \nu_2\sigma) }. 
\end{eqnarray}
>From this ansatz we see that 
\begin{equation}
 t= \frac{ r_+ ( g(\tau) + \nu_2\sigma)  + r_-( f(\tau)+ \nu_1\sigma)}{r_+^2 - r_-^2}, 
\qquad
\phi = \frac{ r_- ( g(\tau) + \nu_2\sigma)  + r_+( f(\tau) + \nu_1\sigma)}{r_+^2 - r_-^2}. 
\end{equation}
The periodicity conditions $t(\tau, \sigma + 2\pi) = t(\tau, \sigma )$ and 
$\phi(\tau, \sigma + 2\pi) = \phi + 2\pi k$ gives rise to the following conditions 
on $\nu_2, \nu_2$. 
\begin{equation}
\label{valnu}
 \nu_1  = r_+ k, \qquad \nu_2 = -r_- k. 
\end{equation}
Now substituting this ansatz in the action  (\ref{coord-act}) we obtain
the following Lagrangian 
\begin{eqnarray}
 L &=& \frac{1}{2} \left( (\dot a)^2 - (\dot f)^2 a^2 - (\dot b)^2 + (\dot g)^2 b^2 +
 \nu_1^2 a^2 - \nu_2^2 b^2 \right. \\ \nonumber
& & \left. -\Lambda( -a^2 + b^2 +1) \right). 
\end{eqnarray}
Again, one can eliminate the constraint by the substitution
\begin{equation}
 a(\tau) = \cosh\gamma(\tau), \qquad b(\tau) = \sinh\gamma(\tau).
\end{equation}
The equations of motion can then be obtained from the Lagrangian
\begin{equation}
 L = \frac{1}{2} \left( - (\dot\gamma)^2  - (\dot f )^2 \cosh^2\gamma + (\dot g)^2
\sinh^2\gamma  
+ \nu_1^1 \cosh^2\gamma - \gamma_2^2 \sinh^2\gamma \right).
\end{equation}
As before we have two constants of motion
given by 
\begin{equation}
\dot f \cosh^2 \gamma = c_1, \qquad \dot g \sinh^2 \gamma = c_2. 
\end{equation}
These constants are related to the charges $E$ and $S$ by the equations in (\ref{eands}). 
We can reduce the equations of motion in $\gamma$ to quadratures using the
Virasoro constraints. This is given by 
\begin{eqnarray}
\label{integral2}
  (\dot\gamma)^2 + \frac{c_1^2}{\cosh^2\gamma} - \frac{c_2^2}{\sinh^2\gamma} +
 \nu_1^2  \cosh^2\gamma - \nu_2^2 \sinh^2 \gamma + (\hat  J^2 +\hat m^2)  &=& 0, \\ \nonumber
 c_1\nu_1 + c_2\nu_2 +\hat  J\hat m &=& 0.
\end{eqnarray}
Note that $\nu_1, \nu_2$ are fixed once the sector label $k$ is given by (\ref{valnu}). 
It is clear that given the constants of motion
 $c_1, c_2, J$ and the sector label $k$ ,  one can determine $\hat m$ 
 from the second of the above equations. 
 The shape of the potential in the first equation shows that  solutions always fall into the horizon. 
This  is because given the constants $c_1, c_2, k, \hat J$, on can always find 
$\gamma_0$ for which 
\begin{equation}
 \frac{c_1^2}{\cosh^2\gamma_0} - \frac{c_2^2}{\sinh^2\gamma_0} +
 \nu_1^2  \cosh^2\gamma_0 - \nu_2^2 \sinh^2 \gamma_0 + ( \hat J^2 +\hat m^2)  = 0, 
\end{equation}
We use this $\gamma_0$ as the initial condition, this of course has $\dot \gamma=0$ and let 
the string evolve and fall into the horizon. 
 The initial condition for the other  co-ordinates are
$t(0, 0) =0$, and  $\phi(0,0)  = 0$. 
With these initial condition, it is certainly possible to integrate the equations of 
motion to obtain the trajectory of the string. We have obtained the trajectory of  $\gamma$ in 
appendix B. 

Without knowing the explicit form of the solution, it is possible to 
obtain the monodromy matrix. We will demonstrate  this below. 
Let us define the matrix
\begin{equation}
  \Omega(x, \sigma ) = P \exp\left[ - \int_0^{\sigma } d\sigma \frac{1}{2} 
\left( \frac{ j_+}{ 1-x} - \frac{j_-}{1+x} \right) \right].
\end{equation}
Note the quasi-momentum  is evaluating the following expression. 
\begin{equation}
\label{wiqua}
 2\cos p(x) = {\rm Tr} (  A_{(k)}  \Omega(x) ).
\end{equation}
$\Omega(x)$ satisfies the differential equation 
\begin{equation}
\label{diffeq}
 \partial_\sigma \Omega( x, \sigma)  = M \Omega (x, \sigma), 
\end{equation}
where the entries of $M$ is given by 
\begin{eqnarray}
M_{11} &=& \frac{1}{1-x^2} \left( x ( c_1+c_2)  +
 \nu_1 \cosh^2 \gamma_0 + \nu_2 \sinh^2 \gamma_0\right), \\ \nonumber
M_{22} &=& - M_{11}, \\ \nonumber
M_{12} &=& \frac{1}{1-x^2}\left[  \left( \frac{\cosh\gamma_0}{\sinh\gamma_0} c_2 + 
\frac{\sinh\gamma_0}{\cosh\gamma_0} c_1 \right) x + ( \nu_1 + \nu_2) \cosh\gamma_0\sinh\gamma_0
\right] e^{( \nu_2 - n_1) \sigma}, 
\\ \nonumber
M_{21} &=& -\frac{1}{1-x^2}\left[ \left( \frac{\cosh\gamma_0}{\sinh\gamma_0} c_2 + 
\frac{\sinh\gamma_0}{\cosh\gamma_0} c_1 \right) x + ( \nu_1 + \nu_2) \cosh\gamma\sinh\gamma_0
\right]e^{- ( \nu_2 - n_1) \sigma}.
\end{eqnarray}
To un-clutter the equations let us define
\begin{eqnarray}
 \tilde A &=& M_{11},  \\ \nonumber
\tilde B &=&  \frac{1}{1-x^2}\left[ \left( \frac{\cosh\gamma_0}{\sinh\gamma_0} c_2 + 
\frac{\sinh\gamma_0}{\cosh\gamma_0} c_1 \right) x + ( \nu_1 + \nu_2) \cosh\gamma\sinh\gamma_0
\right].
\end{eqnarray}
It is easy to integrate the differential equation in (\ref{diffeq}) with the initial 
conditions $\Omega(x, 0) = 1$.  This results in 
\begin{eqnarray}
 & & \Omega(x, \sigma) =\\ \nonumber
 & & \frac{1}{\lambda_+ - \lambda_-} \left(
\begin{array}{cc}
 -( \tilde A  + \lambda_-) e^{\lambda_+\sigma} + 
(\tilde A + \lambda_+) e^{\lambda_- \sigma}   & 
 \tilde B ( - e^{\lambda_+\sigma} + e^{\lambda_-\sigma}) \\
\tilde B (- e^{-\lambda_+ \sigma} +  e^{-\lambda_-\sigma}) 
&  -( \tilde A  + \lambda_-) e^{- \lambda_+\sigma} + 
(\tilde A + \lambda_+) e^{- \lambda_- \sigma} 
\end{array}
\right), 
\end{eqnarray}
where $\lambda_\pm$ are given by
\begin{equation}
 \lambda_{\pm} = \frac{1}{2} \left( \nu_2-\nu_1 \pm 
\sqrt{ ( \nu_2- \nu_1)^2 - 4 ( \tilde B^2 - \tilde A^2 - ( \nu_2-\nu_1) \tilde A)  } \right). 
\end{equation}
We can simplify the terms in the square root using the Virasoro constraints to 
obtain
\begin{eqnarray}
\nonumber 2 \lambda_{\pm} &=& \nu_2-\nu_1 \pm \tilde  D,  \\ \label{defd}
\tilde D & = &   \left\{ ( \nu_2- \nu_1)^2  -
  \frac{4}{( 1-x^2)^2} \left( 
 \frac{\hat J^2}{4\pi^2\lambda^2}  + m^2 + 2 x \frac{ \hat J\hat m }{2\pi \lambda}  \right) 
\right. \\ \nonumber
& & \left.  + \frac{4}{ 1-x^2} \left( 
\frac{c_2^2}{\sinh^2\gamma_0} - \frac{c_1^2}{\cosh^2\gamma_0} \right)
+ 4 ( \nu_2 - \nu_1) \tilde A \right\}^{1/2}. 
\end{eqnarray}
We can now find the quasi-momentum by evaluating (\ref{wiqua}),
after substitution  for $\Omega$  and using (\ref{valnu}) we obtain 
\begin{equation}
 \cos p(x) =  \cosh \pi  \sqrt{\tilde D}. 
\end{equation}
which implies 
\begin{equation}
\label{quasi-ans}
 p = i \pi \sqrt{\tilde D}. 
\end{equation}
Thus we have explicitly solved for the quasi-momentum for these classical solutions which represent
winding strings. 

We can now verify the behaviour of the quasi-momentum when the spectral 
parameter $x$ is taken to be near $\pm1$, $0$  and $\infty$. 
When $x\rightarrow \pm1$, it can be seen from  the expression for $\tilde D$ in 
(\ref{defd}), that the behaviour given in (\ref{k1})  is reproduced.
Now as $x\rightarrow \infty$ we see that the leading  contributions 
comes form the terms $(\nu_2 - \nu_1)^2$ and  the term proportional 
to $c_1+c_2$ in $\tilde A$  in the expression for $\tilde D$. 
 Going through the analysis we find the 
following behaviour as $x\rightarrow \infty$.
\begin{eqnarray}
\label{solinfw}
 p(x) &=&   i \pi k ( r_++ r_-)  +  i \frac{ 2\pi }{x} (c_1+ c_2)+ \cdots  , \qquad x\rightarrow \infty, \\ \nonumber
&=&  i \pi k ( r_++ r_-)  -  i \frac{ E-S}{ \lambda x ( r_+ + r_-) } + \cdots . 
\end{eqnarray}
where we have used (\ref{eands}) to obtain the second line in the above equation. 
Comparing (\ref{solinfw})  with equation (\ref{beinf}) obtained from general considerations
we see that they are in agreement. 
When $x\rightarrow 0$, the leading terms in $\tilde D$ are given by
\begin{equation}
 \tilde D \sim ( \nu_1 - \nu_2)^2 + 4 \nu_1\nu_2  -4 x ( c_2 - c_1) ( \nu_1 + \nu_2). 
\end{equation}
From this behaviour we see $p$ behaves as
\begin{eqnarray}
 p(x) &=& i \pi( r_+ - r_-) k + i 2\pi x( c_1 - c_2) + \cdots , \qquad x\rightarrow 0 , \\ \nonumber
 &=& i \pi ( r_+ - r_-) k  + i x \frac{E+S}{\lambda ( r_+ - r_-) }+ \cdots . 
\end{eqnarray}
We see that the above equation 
is in agreement with (\ref{pzero}) obtained from general considerations with $ m =0$.
Finally from (\ref{quasi-ans}) we see that just above and below  the branch cuts  the quasi-momentum just 
flips sign, thus the condition (\ref{jumcon}) is obeyed with $n_l=0$. 

Since we have the explicit expression for the quasi-momentum given by (\ref{quasi-ans}) we can 
obtain the non-local charges ${\cal C}$ and ${\cal Q}$ defined in (\ref{nonlocal1}) and 
(\ref{nonlocal2}) respectively. 
Expanding the quasi-momentum to $O(1/x^2)$ as $x\rightarrow\infty$ we obtain 
\begin{eqnarray}
{\cal C} =
2\pi^2(c_1+c_2)^2\cosh \pi k ( r_+ + r_-)+\frac{2\pi h}{\nu_2-\nu_1} \sinh \pi k ( r_+ + r_-), 
\end{eqnarray}
where
\be
h=\frac{c_2^2}{\sinh^2\gamma_0}-\frac{c_1^2}{\cosh^2\gamma_0}+
\nu_2^2 \sinh^2\gamma_0 -\nu_1^2\cosh^2\gamma_0 +\nu_1\nu_2.. 
\ee
Similarly from the 
 $O(x^2)$ terms in the $x\rightarrow 0$ expansion of the quasi-momentum we can 
read out the charge ${\cal Q}$ which is given by
\begin{eqnarray}
{\cal Q} = 2\pi^2(c_2-c_1)^2\cosh \pi k ( r_+ + r_-) 
-\frac{2 \pi \delta}{\nu_2 + \nu_1}  \sinh{\pi k ( r_+ + r_-)}
\end{eqnarray}
where
\be
\delta=\frac{c_2^2}{\sinh^2\gamma_0}-\frac{c_1^2}{\cosh^2\gamma_0}+
\nu_2^2 \cosh^2\gamma_0 -\nu_1^2 \sinh^2\gamma_0+\nu_1\nu_2. 
\ee

Using this example of winding strings 
we have verified the conditions on the quasi-momentum found in the previous section and also demonstrate
that this solution corresponds to a density localized on the branch cuts of the quasi-momentum $p(x)$ given 
in (\ref{quasi-ans}).

\subsection{BMN and magnon like states}

In  section 3.2  we have listed out the conditions satisfied the by resolvent in 
terms of  the density $\rho(x)$. These are given in (\ref{den1}), (\ref{den2}), (\ref{den3}) and 
(\ref{den4}).  
 In this section we find two simple solutions to these
equations which resemble the BMN states and magnon states found for
classical strings propagating on $R\times S^3$ 

\vspace{.5cm}
\noindent
{\bf BMN like states}

These are solutions to the resolvent obtained when the 
branch cuts shrink to delta functions.  For simplicity in illustrating these solutions 
let us look at the situation when the winding number on $S^1$ given by $\hat m$ vanishes and work in the 
sector $k\neq 0$. 
Let these delta functions be localized at 
$x_s$ with strength $S_s$. Then from (\ref{den4}), these positions satisfy
\begin{equation}
 \frac{1}{x_s} = - \frac{\hat J}{2\pi \lambda ( n_s + i k(r_+ + r_-) )} 
\left( 1 - \sqrt{ 1+ \frac{4\pi^2 \lambda^2}{\hat J^2}(  n_s + i k( r_+ + r_-) ) ^2} \right), 
\end{equation}
The equations (\ref{den1}), ( \ref{den2}) and ( \ref{den3}) reduces to 
\begin{eqnarray}
\label{bmneq}
 \sum_s S_s  &=& -\frac{\hat J}{\lambda} + i \frac{ E-S}{\lambda( r_+ + r_-)}, \\ \nonumber
-\sum_s \frac{S_s}{x_s} &=& 0,  \\ \nonumber
-\sum_s \frac{S_s}{x_s^2} &=& \frac{\hat J}{\lambda} + i \frac{E+S}{\lambda( r_+-r_-)}. 
\end{eqnarray}
We have set $ m=0$ since in this approximation the filling fractions are 
small, we  must also have $r_-= 0$ for this approximation 
to be valid.  

Following \cite{Kazakov:2004qf}, we parametrize the strength of the delta functions as 
\begin{equation}
 S_s = \frac{N_s J}{2\pi \lambda}
 \left( 1 + \sqrt{ 1+ \frac{4\pi^2 \lambda^2}{\hat J^2}(  n_s + i k  r_+  ) ^2} \right).
\end{equation}
Just as in \cite{Kazakov:2004qf} we can associate $N_s$ to be the occupation numbers. 
With this parametrization (\ref{den2}) reduces to 
\begin{equation}
 \sum_s ( n_s + i k r_+  )N_s =  0. 
\end{equation}
Since this is a complex equation and  $N_s$ are taken to be real, it results in the 
following   two  real equations 
\begin{equation}
 \sum_s n_s N_s =0, \qquad  \sum _s N_s =0. 
\end{equation}
Now adding the first  and the third equations in (\ref{bmneq}) we obtain
\begin{equation}
\frac{\hat J}{\lambda} \sum N_s = - i \frac{ E-S}{\lambda( r_+ + r_-)} + i \frac{E+S}{\lambda( r_+-r_-)} =0. 
\end{equation}
Finally from the third equation in (\ref{bmneq}) we obtain
\begin{equation}
\label{bmndisp}
  \frac{\hat J}{\lambda} + i \frac{E+S}{\lambda( r_+-r_-)} = \sum_s
\frac{N_s \hat J}{2\pi \lambda}  \sqrt{ 1+ \frac{4\pi^2 \lambda^2}{\hat J^2}(  n_s + i k  r_+  ) ^2}. 
\end{equation}
Here we have used the fact that $\sum_s N_s =0$.  
Matching the real and imaginary parts of the above dispersion relation we obtain 
the quantum numbers carried by the states in terms of the occupation numbers $N_s$ and the frequency
$n_s$.  We call these states BMN-like states, since they have a similar dispersion relation.

\vspace{.5cm}
\noindent
{\bf Magnon-like states}
\vspace{.5cm}

Magnon-like solutions can be obtained from the equations  (\ref{den1}), ( \ref{den2}) and (\ref{den3}) 
by following the same procedure as in the case of the sigma model on $S^3$ which was
done in \cite{Minahan:2006bd}.  
The procedure consists of assuming the momentum of the magnon is given by the 
LHS of (\ref{den2}) can be any number $p$.
The density is $\rho(x)$ is  
 that the density is $\rho(x)$ is constant along a contour in the 
$x$-plane given by $i\rho =1$ say between $x_1$ and $x_2$
\footnote{For the case of magnons on $S^3$ the contour had to be a line 
between two complex conjugate points to ensure that the momentum
of the magnon is real.}. 
There are no branch cuts for the density function,  therefore the equation (\ref{den4})
does not play a role. 
 Substituting this ansatz in (\ref{den1}), (\ref{den2}) and (\ref{den3}) 
we obtain the following equations 
\begin{eqnarray}
 -i( x_1 - x_2) &=& - \frac{\hat J}{\lambda} - \frac{i}{\lambda} \frac{ E-S}{r_+ + r_-}, \\ \nonumber
- i \ln\frac{x_1}{x_2} &=& p,  \\ \nonumber
-i \left( \frac{1}{x_1} - \frac{1}{x_2} \right) &=& \frac{\hat J}{\lambda} + 
\frac{i}{\lambda} \frac{E+S}{r_+ - r_-}. 
\end{eqnarray}
As mentioned earlier   we have replaced 
the combination $2\pi( \tilde m + m + i kr_-)$ as the momentum. 
The momentum in this case can be complex and thus $x_1$ and $x_2$ need not be 
at two complex conjugate points.  
We can now solve for the $x_1, x_2$ and obtain the following dispersion 
relations for the quantum numbers $E, S$. 
\begin{equation}
Q_+ - 2\hat Ji= \sqrt{ Q_-^2 - 16\lambda^2 \sin^2 \frac{p}{2} }, 
\end{equation}
where
\begin{eqnarray}
\label{magdisp}
Q_+ =   \frac{ E+ S}{r_+ - r_-} + \frac{E-S}{r_+ + r_-} , \qquad
Q_-  =  \frac{ E+ S}{r_+ - r_-} -\frac{E-S}{r_+ + r_-} . 
\end{eqnarray}

Both equations (\ref{bmndisp}) and (\ref{magdisp}) are 
resemble the dispersion relations of the  quasi-normal modes found by studying 
the wave equations in the BTZ black hole \cite{Birmingham:2001pj} due to the presence of the imaginary 
parts. It will be interesting to find explicit classical solutions obeying these dispersion 
relations.

\subsection{Relation with the $SL(2, R)$ spin chain}

In this section we show 
that the system of equations (\ref{den1}), (\ref{den2}), (\ref{den3}) and
(\ref{den4}) which characterize classical solutions in the twisted sector,
 of the BTZ sigma model can be obtained from the 
the continuum limit  of a twisted version of the $SL(2, R)$ spin chain. 
to the properties of the resolvent of the
sigma model. 
The system we will consider is the twisted long range $SL(2, R)$ chain. 
The Bethe equations of this model are given by 
\begin{equation}
\label{sbethe1}
\left( \frac{ x( u_k + \frac{i}{2} )}{ x(u_k - \frac{i}{2} )} 
\right)^{L}  \exp(2i  \pi k ( \tilde r_+  + \tilde r_-) ) 
= \prod_{j= 1, j\neq k}^{M } \frac{ u_k - u_j - i}{ u_k - u_j + i}, 
\end{equation}
where $x$'s label the Bethe roots. 
Note that the important difference between these Bethe equations and 
the ones for the $SU(2)$ spin chain given in (\ref{bethe1}) is the
inversion of the RHS for the case of the $SL(2, R)$ chain. 
As in the earlier case, $x$ as a function of $u$ is given by
the equation (\ref{chgvarxu}).  
The cylicity constraint in this case is given by
\begin{equation}
\label{sbethe2}
\prod_{k=1}^{M}\left( \frac{ x( u_k + \frac{i}{2} )}{ x(u_k - \frac{i}{2} )} 
\right) =  \exp(2\pi i  k \tilde r_-). 
\end{equation}
The energy of this spin chain is given by 
\begin{equation}
\label{sbethe3}
 D = 2g^2 \sum_{i=1}^{M} \left( \frac{i}{ x(u + \frac{i}{2} ) } - \frac{i}{x(u -\frac{i}{2} ) } \right). 
\end{equation}
These set of equations defines the twisted version of the $SL(2, R)$ chain.

Let us first obtain the continuum limit of these equations by  performing  the scaling
$u_k \rightarrow L u_k$. Then taking the logarithm of the Bethe  equations (\ref{sbethe1}) we obtain
\begin{equation}
L \ln\left( \frac{ 1- \frac{i}{2 L u_k} }{ 1 + \frac{i}{2 L u_k} } \right)
- 2\pi i k (\tilde  r_+ + \tilde r_-) + 2\pi i n 
= \sum_{j\neq k} \ln
 \left(\frac{ 1+ \frac{i}{L(u_k - u_j)} }{ 1- \frac{i}{ L( u_k -u_j)}} \right). 
 \end{equation}
 Approximating the sum by an integral and expanding to $O(1/L)$  we obtain
 \begin{equation}
\label{cbethe1}
 -\frac{1}{u} - 2\pi k ( \tilde r_+ +\tilde  r_-)  +  2\pi n =   2 \pint dv \frac{\rho( u)}{ u-v}, 
\end{equation}
where we have introduces a density for the spin. The density statisfies 
the normalization condition 
\begin{equation}
\label{cbethe2}
\int du \tilde \rho(u) = \frac{M}{L}. 
\end{equation}
The cyclicity constraint given in (\ref{sbethe2}) reduces to 
\begin{equation}
\label{cbethe3}
\int du \frac{\tilde \rho(u)}{u} = -2\pi ( \tilde m-  k \tilde r_-) . 
\end{equation}
Finally the energy of the spin chain given in (\ref{sbethe3})  becomes
\begin{equation}
\label{cbethe4}
 D =  \frac{2g^2}{L} \int du \frac{\tilde \rho(u) }{ u^2} + O(g^4). 
\end{equation}

To relate these equations to that of the integral  equations satisfied by the 
density in  the  sigma model case, 
we need to first identify the relation between the variable $u$ of the spin chain 
with the spectral parameter of the sigmal model. 
This is given by 
\begin{equation}
u = x + \frac{g^{\prime 2}}{x} , \qquad {\hbox{ where}} \quad g' = \frac{g}{L}. 
\end{equation}
Then the normalization condition for the density (\ref{cbethe2})  becomes
\begin{equation}
\label{norm-int}
\int dx ( 1 - \frac{g^{\prime 2} }{x^2} ) \tilde \rho ( u(x)) = \frac{M}{L}. 
\end{equation}
Again using the relation in (\ref{redefb}) we can rewrite the Bethe equations (\ref{cbethe1} 
as 
\begin{equation}
\label{beth2}
-  \frac{x}{x^2 - g^{\prime 2} } - 2\pi k ( \tilde r_+ + \tilde r_-) 
 + 2\pi n = 2\pint  dy \left( 1- \frac{g^{\prime 2}}{y^2} \right) 
\frac{\tilde\rho( u(x)) }{ ( x -y)( 1- \frac{g^{\prime 2} }{xy} )}. 
\end{equation}
In the leading order in $g$ this equation reduces to 
\begin{equation}
\label{beth1}
 -\frac{1}{x} - 2\pi k ( \tilde r_+ + \tilde r_-)  + 2\pi n = 2 \pint dy \frac{\tilde\rho( u(x)) }{ ( x -y)}. 
\end{equation}
The cyclicity contraint also reduces to 
\begin{equation}
\label{beth22}
 \int \frac{\tilde \rho(x)}{x} = - 2\pi (\hat m - k \tilde r_+), 
\end{equation}
where we have again used (\ref{redefb}). 
Finally  the energy of the spin chain to the leading order is given by 
\begin{equation}
\label{beth3}
 \frac{D }{L} = 2g^{\prime 2}  \int dx \frac{\tilde \rho(x)}{x^2}. 
\end{equation}

We can now compare the Bethe equations of the spin chain to that of the  sigma model. 
First consider the sum of the equations  given in (\ref{den1}) and 
(\ref{den3}) we see that we obtain
\begin{equation}
\label{norm-sig}
\int d\xi \left( 1 - \frac{1}{\xi^2} \right) \rho(\xi) = \frac{i}{2} 
\left( -\frac{ E- S}{\lambda( r_+ + r_-)} 
+ \frac{ E+S}{\lambda( r_+ - r_-)} \right). 
 \end{equation}
To make the Bethe equations resemble that of the jump condition  across branch cuts satisfied by 
of the resolvent of the sigma model  we first scale the spectral parameter of the 
sigma model by 
\begin{equation}
x\rightarrow  \frac{2\hat J}{\lambda} x
\end{equation}
Substituting this scaling  in (\ref{den4})  we get 
\begin{eqnarray}
\label{sig1}
2\pint d\xi \frac{\rho(\xi)}{x-\xi} &=& - 
\frac{ x + \frac{\pi\lambda^2}{\hat J^2}\hat m }{x^2 - \frac{\lambda^2}{4\hat J^2} } 
+ 2\pi n - 2\pi i k ( r_+ + r_-) , \\ \nonumber
&=& - \frac{1}{x} + 2\pi n - 2\pi i k ( r_+ + r_-)  + O(\lambda^2), 
 \end{eqnarray}
 while the equation in (\ref{norm-sig}) reduces to 
 \begin{equation}
\label{norm-sig2}
\int d\xi \left( 1 - \frac{\lambda^2}{4 \hat J^2 \xi^2} \right) \rho(\xi) = - \frac{i}{4\hat J} 
\left( \frac{ E- S}{( r_+ + r_-)} 
- \frac{ E+S}{( r_+ - r_-)} \right). 
 \end{equation}
The equation (\ref{den2}) of the sigma model remains invariant under  the scaling and is given by 
\begin{equation}
\label{sig2}
 \int dx \frac{\rho(x)}{x} = - 2\pi( \hat m + m - i kr_-) . 
\end{equation}
 Finally the equation in (\ref{den3}) becomes 
\begin{equation}
\label{sig3}
- \frac{\lambda^2}{2 \hat J^2} \int dx  \frac{\rho(x)}{x^2} = 1  + i \frac{E+ S}{ \hat J(r_+ - r_-)} . 
\end{equation}

We now see that on identification  the equation (\ref{beth1}) 
(\ref{beth22}) and (\ref{beth3}) and (\ref{norm-int}) 
 of the spin chain are the same as (\ref{sig1}), (\ref{sig2}),  (\ref{sig3}) and (\ref{norm-sig2})  of the 
sigma model on the following identifications.
\begin{eqnarray}
 & & g' = \frac{g}{L}   = \frac{\lambda}{2\hat J},   \qquad
\tilde\rho(u(x)) =  \rho(x), \\ \nonumber  
& & i r_+ \rightarrow \tilde r_+,  \qquad i r_- \rightarrow  \tilde r_-,  \qquad \hat m + m \rightarrow \tilde m \\ \nonumber
& &  \frac{1}{4 \hat J} \left( \frac{ E- S}{( \tilde r_+ + \tilde r_-)} -
 \frac{ E+S}{( \tilde r_+ - \tilde r_-)} \right) = \frac{M}{L}, \\ \nonumber
& & \frac{D}{L} =- 1 + \frac{E+ S}{\hat J ( \tilde r_+ - \tilde r_-)}. 
\end{eqnarray}
Note that this identification involves the analytical continuation of the 
parameters $r_+, r_-$. 
This completes our proof that at one loop,  the Bethe equations of the 
twisted $SL(2, R)$ spin chain agrees with that of the finite gap equations of the 
sigma model.

\section{Conclusions}

We have shown that the sigma model on BTZ $\times S^1$ is integrable
using the fact that it is locally $AdS_3$. 
We construct the monodromy matrix
of the flat connection  and studied the general properties of the 
quasi-momentum. 
We have obtained integral equations which constrains the quasi-momentum 
 and shown that classical solutions correspond to a 
density function on the spectral plane. These integral equations   have been
shown to agree with the continuum limit of a twisted version of the $SL(2, R)$ spin chain at one 
loop. 
For two class of solutions, the geodesics and the winding strings
we have evaluated the corresponding quasi-momentum  explicitly. 
Using this we verify that the its properties agree with that obtained
by general consideration.  Geodesics correspond to solutions 
with zero density in the spectral plane. 

We solved the integral equations for the denisty funtion for the   BMN like and 
magnon like solutions and derived their dispersion relations. 
It will be interesting to construct explicit solutions 
of the sigma model corresponding to these BMN like and magnon like solutions.  
Though the BTZ black hole is a well studied, 
the propagation of strings in this background has not been studied
in great  detail. It will be interesting to  
find the allowed string spectrum in this background.  A step in this direction will be 
to find more classical solutions. 
Our investigations indicate that  integrability will be a useful structure to 
find and organize the spectrum. 
This will have implications for the dual conformal theory corresponding 
to the BTZ background just as quasi-normal modes of the scalar field correspond 
to poles in the two point function of the dual operator.

\acknowledgments

We wish to thank Rajesh Gopakumar,  Gautam Mandal, Shiraz Minwalla, 
Ashoke Sen and Spenta  Wadia  for useful discussions and comments. 
We thank the International Centre of Theoretical Sciences (ICTS), 
 of the TIFR for  organizing a stimulating discussion meeting at IISc, Bangalore 
 during  which the idea for this work originated. 
The work of  J.R.D is partially supported by 
the Ramanujan fellowship DST-SR/S2/RJN-59/2009, the work of A.S is supported by  a 
CSIR fellowship (File no: 09/079(2372)/2010-EMR-I). 

 \appendix

\section{The case of the extremal black hole}

For completeness we wish to repeat the analysis of 
the quasi-momentum done in main text of the paper  for the extremal BTZ black hole. 
The identification used in the section 3 to obtain  the non-extremal black hole 
from the $AdS_3$ hyperboloid cannot be used for the 
extremal case \cite{Banados:1992gq}.
The metric for the extremal 
BTZ black hole is given by 
\begin{equation}
\label{extbtz}
 ds^2 = - \frac{ (r^2 - r_0^2)^2}{r^2 }dt^2 + \frac{r^2}{(r^2 - r_0^2)^2} dr^2 
+ r^2 ( d\phi + \frac{r_0^2}{r^2}dt ) ^2 , 
\end{equation}
where $r_0$ is the location of the horizon. 
We can show that the matrix is locally $AdS_3$ by the following change of 
coordinates. Let us define
\be
w^+=\frac{e^{2r_0(\phi +t)}}{2r_0}, \qquad w^-=\phi-t-\frac{r_0}{r^2-r_0^2}, \qquad z=\frac{e^{r_0(\phi +t)}}{\sqrt{r^2-r_0^2}}, 
\ee
then the metric in (\ref{extbtz}) reduces to the Poincare metric given by 
\begin{equation}
 ds^2=\frac{dz^2+dw^+ dw^-}{z^2}.
\end{equation}
However since $\phi$ is a periodic coordinate, the Poincare metric is 
subject to  indentifications.  To write this as a group action, we identify the 
$SL(2, R)$ group element as
\begin{equation}
 g = \left( \begin{array}{cc}
\frac{1}{z} & \frac{w^-}{z} \\
\frac{w^+}{z} & \frac{z^2+w^-w^+}{z}
\end{array} \right).
\end{equation}
Then the identification 
\begin{equation}
 \phi \sim \phi + 2\pi,\qquad
{\mbox{acts as}} \qquad
 g \sim \tilde A_{(1)} g A_{(1)},
\end{equation}
where
\begin{eqnarray}
 \tilde{A}_{(k)} = \left( \begin{array}{cc}
e^{-2\pi kr_0} & 0 \\
0 & e^{2\pi kr_0}
\end{array} \right),  \qquad
A_{(k)} = \left( \begin{array}{cc}
1 & 2\pi k \\
0 & 1
\end{array} \right).
\end{eqnarray}
Thus the monodromy matrix for the extremal BTZ background is 
given by (\ref{fullmono}) but with $A_k$ given in the above equation. 
The conserved charges $E$ and $S$ in this case are given by 
\begin{eqnarray}
E+S=-\lambda \left[ \int_0^{2\pi} d\sigma \left(r_0 {\rm Tr} (\partial_0 g g^{-1} \sigma^3)\right)\right],  \\
E-S=-\frac{\lambda}{2} \left[ \int_0^{2\pi} d\sigma \left({\rm Tr}(g^{-1} \partial_0 g\sigma_+)\right)\right], 
\end{eqnarray}
where $\sigma_+ = \sigma^1 + i \sigma^2$. 
The behaviour of the quasi-momentum as $x\rightarrow \pm 1$ is same as before and is 
given by (\ref{k1}). 
We now examine its behaviour as $x\rightarrow \infty$. 
\begin{eqnarray}
2 \cos p(x) &=& {\rm Tr} 
\left[ A_k P \exp
\left( \int_0^{2\pi} ( d\sigma \frac{j_0}{x} + \frac{j_1}{x^2} \cdots) \right)   \right],  \\ \nonumber
&=& {\rm Tr} \left((1+\pi k \sigma_+)(1+\int_0^{2\pi}d\sigma \frac{j_0}{x} + \frac{j_1}{x^2} \cdots)\right),  \\ \nonumber
&=& 2+\frac{\pi k}{x}{\rm Tr} \left( \int_0^{2\pi}\sigma_+j_0d\sigma \right) + \cdots,  \\ \nonumber
&=&2-\frac{\pi k(E-S)}{x\lambda. }. \\  
\end{eqnarray}
Therefore the behaviour of the quasi-momentum is  given by
\be
p(x)\sim \sqrt{\frac{\pi k(E-S)}{x\lambda}} + \cdots , \qquad x\rightarrow \infty. 
\ee
Note that the quasi-momentum seems to have a square root branch cut at $\infty$. 
At $x\rightarrow 0$ we have
\begin{eqnarray}
2 \cos p(x) &=&{\rm Tr}( \tilde A_k^{-1}( 1 - x  \int_0^{2\pi} d\sigma  \partial_\tau g g^{-1}  + \cdots ) ), \\ \nonumber
&=&{\rm Tr}\left[  ( \cosh 2\pi k r_0 + \sigma^3 \sinh 2\pi k r_0  ) 
( 1 - x  \int_0^{2\pi} d\sigma \partial_\tau g g^{-1}  + \cdots )\right], \\ \nonumber
&=&2 \cosh 2\pi k r_0 + x \sinh2 \pi k r_0 \frac{E + S}{2\lambda r_0}.
\\ \nonumber\end{eqnarray}
The final result for the behaviour of the 
quasi-momentum at $x\rightarrow 0$ as 
\begin{equation}
p ( x) \sim  2\pi \tilde m + i 2\pi k r_0 + i \frac{x}{4}   \frac{E + S}{\lambda r_0 }.
\end{equation}
The fact that the quasi-momentum has a square root branch cut as $x\rightarrow\infty$ possibly
implies that to construct
the resolvent one has to consider the double cover of the spectral plane for this case. 
It will be interesting to investigate this further. 

\section{Trajectories}

In this section we integrate the equations (\ref{integral1}) and (\ref{integral2}) for both the
geodesics and the winding strings. Note that the (\ref{integral2}) reduces to 
(\ref{integral1}) on setting $\nu_1=\nu_2 =0$, therefore
for generality let us examine (\ref{integral2}). 
Solving for $\gamma$ as the function of the world sheet time we obtain the 
following integral
\be
\label{integral3}
\int d\tau=\frac{1}{2}\int \frac{dx}{\sqrt{Ax^3+Bx^2+Cx+D}}, 
\ee
upon substituting 
\begin{eqnarray}
 x=\sinh^2{\gamma}
\end{eqnarray}
The constants $A, B, C, D$ are given by 
and
\be
A=\nu^2_2-\nu^2_1, \quad B=\nu^2_2-2\nu^2_1, \quad C=c^2_2-c^2_1-\nu^2_1, \quad 
D=c^2_2-(\hat J^2+\hat m^2). 
\ee

\vspace{.5cm}
\noindent
{\bf Geodesics }

We first rewrite the equation
 (\ref{potential}) in terms of the BTZ  variables and parameter $r,M,j$ using the relations
\begin{eqnarray}
 c_2^2-c_1^2=\frac{E^2-S^2}{r_+^2-r_-^2}, \quad r_{\pm}^2=\frac{M}{2}\pm \frac{\sqrt{M^2-j^2}}{2}, \quad \sinh ^2{\gamma}=\frac{r^2-r_+^2}{r^2_+-r^2_-}
\end{eqnarray} 
(\ref{rmj}) and (\ref{eands}). 
Note that we have set $\lambda =2\pi$. 
 We then  obtain the following equation for particle trajectory
\be
\label{pargeo}
r^2\dot r^2=-\hat J^2(r^4-Mr^2+\frac{j^2}{4})+(E^2-S^2)r^2+S^2M+ESj
\ee
Now this matches exactly with equation (8) of \cite{Cruz:1994ir} provided we do the following identifications
\be
S=-L, \quad \hat J=m
\ee
where $m$ is the particle mass and $L$ is the conserved angular momentum in the paper \cite{Cruz:1994ir}. Hence the geodesics given in the paper will also follow.

\vspace{.5cm}
\noindent
{\bf Winding strings  }

The integral (\ref{integral3}) can be solved analytically to give an answer involving 
Here will will just outline the procedure for solving the integral. 
Let us denote the roots of the equations 
\be
Ax^3+Bx^2+Cx+D= A( x-x_1)( x-x_2) (x-x_3) =0, 
\ee
as $x_1,x_2,x_3$. Then the integral can be written as 
\be
\int d\tau= \frac{1}{2\sqrt{A} }\int \frac{dx}{\sqrt{(x-x_1)(x-x_2)(x-x_3)}}.
\ee
After the substitution 
\be
z=\sqrt{\frac{x-x_1}{x_2-x_1}}, 
\ee
the integral reduces to 
\be
\int d\tau=\frac{1}{\sqrt{(x_3-x_1) A }}\int \frac{dz}{\sqrt{(1-z^2)(1-k^2z^2)}}, 
\qquad k=\sqrt{\frac{x_2-x_1}{x_3-x_1}}. 
\ee
which can be integrated using Jacobi elliptic functions. 
Thus we have the following solution for $\gamma(\tau)$
\be
\sinh^2 \gamma - x_1 = ( x_2-x_1) {\rm  sn^2 }( \sqrt{ A( x_1 - x_3)}( \tau + c)  ).
\ee
In passing we mention there are another  case  in which the integral 
reduces to simpler function. The case in which the parameters are chosen so that 
$C=D=0$ the integral can be done in terms of hyperbolic functions.

\providecommand{\href}[2]{#2}\begingroup\raggedright\endgroup

\end{document}